\documentclass[12pt]{article}
\pdfoutput=1

\usepackage{cite}    % Handles sorting and compressing numeric ranges

\usepackage{physics}
\usepackage{booktabs}
\usepackage{amssymb}
\usepackage{mwe}
\usepackage{amsmath}
\usepackage{graphicx}
\usepackage{caption}
\usepackage{subcaption}
\usepackage{tensor}
\usepackage{setspace}
\usepackage{titling} % for abstract
\usepackage{lipsum} % abstract filler
\usepackage[letterpaper, total={6.5in,8in}]{geometry}%,total={6in,8in}]{geometry}
\usepackage{hyperref}
\usepackage{authblk}
\usepackage{upgreek}
\usepackage{color}

\hypersetup{
    colorlinks=true,
    linkcolor=blue,
    filecolor=magenta,      
    urlcolor=cyan,
citecolor=cyan,
    pdftitle={Black hole sensitivity in Einstein--\ae ther theory}
    }

\def\sectionautorefname~#1\null{Section (#1)\null}
\def\subsectionautorefname~#1\null{Section (#1)\null}
\def\equationautorefname~#1\null{(#1)\null}
\def\figureautorefname~#1\null{(Figure #1)\null}

\def\ba{\begin{eqnarray}}
\def\ea{\end{eqnarray}}
\def\be{\begin{equation}}
\def\ee{\end{equation}}

\def\d{\partial}

\renewcommand{\l}{\lambda}

\renewcommand{\b}{\beta}

\newcommand{\vk}{\varkappa}

\newcommand{\vf}{\varphi}

\renewcommand{\a}{\alpha}
\renewcommand{\b}{\beta}

\newcommand\bseq{\begin{subequations}}
\newcommand\eseq{\end{subequations}}

\def\mc#1{\mathcal{ #1 }}

\numberwithin{equation}{section}

\allowdisplaybreaks[4]

\title{{\bf \Large Black hole sensitivity in Einstein--\ae ther theory}}
\author[1,2]{Morgan Kidnew\footnote{kidnewm@mcmaster.ca}}
\author[1]{Andrew Kovachik\footnote{kovachia@mcmaster.ca}}
\author[1,3]{Sergey Sibiryakov\footnote{ssibiryakov@perimeterinstitute.ca}}
\affil[1]{Department of Physics and Astronomy, McMaster University, 
Hamilton, ON L8S 4M1, Canada}
\affil[2]{Department of Physics, University of Guelph, Guelph, ON N1G 2W1, Canada}
\affil[3]{Perimeter Institute for Theoretical Physics, Waterloo, Ontario, N2L 2Y5, Canada}
\begin{document}
\maketitle{}

\begin{abstract}
We numerically find the solutions for slowly moving Schwarzschild black holes in Ein\-stein--\ae ther theory. We use these solutions to compute the sensitivity parameter which describes the coupling of the black hole to the aether field within point-particle effective theory. A simple analytic approximation for the dependence of the sensitivity on the Einstein--\ae ther parameters is derived.
\end{abstract}

\thispagestyle{empty}

\newpage
\setcounter{tocdepth}{2}

{\hypersetup{hidelinks}
\tableofcontents
}

\section{Introduction}

The advent of gravitational wave (GW) astronomy \cite{LIGOScientific:2016aoc} with the rapidly growing number of observed binary mergers \cite{LIGOScientific:2026sit} has opened the possibility to test the laws of gravity in new dynamical regimes \cite{Yunes:2025xwp}. The upcoming generation of GW detectors is expected to push the precision of these tests to unprecedented level \cite{Barausse_2016,Berry:2019wgg}. This calls for development of accurate theoretical methods for modeling the gravitational waveforms in general relativity (GR), as well as its extensions~\cite{LISAConsortiumWaveformWorkingGroup:2023arg}.

One of the cornerstones of GR is the local Lorentz invariance (LI). It has been suggested, however, that LI may not be an exact symmetry of gravity at the quantum level; see \cite{Liberati:2013xla} for review. A concrete proposal was made by 
Ho\v rava \cite{Horava:2009uw}. If LI is not a fundamental symmetry of nature at high energies, one can construct power-counting renormalizable extensions of GR \cite{Mukohyama:2010xz,Sotiriou:2010wn,Blas:2014aca,Herrero-Valea:2023zex,Barvinsky:2023mrv}. One of these extensions (so-called `projectable' model) has been shown to be renormalizable in a rigorous sense \cite{Barvinsky:2015kil,Barvinsky:2017zlx} and possess renormalization group flow with asymptotically free ultraviolet fixed points \cite{Barvinsky:2017kob,Barvinsky:2019rwn,Barvinsky:2021ubv,Radkovski:2023cew,Barvinsky:2023uir,Barvinsky:2024svc}. Thus, it is a complete quantum theory. Low-energy viability of the projectable model is, however, unclear due to instability and strong coupling issues (see \cite{Mukohyama:2026moc} and references therein). By contrast, another, `non-projectable' extension has good low-energy properies, reducing at the lowest-derivative level to a scalar-tensor theory. The scalar describes a foliation of the spacetime by preferred time surfaces \cite{Germani:2009yt,Blas:2009qj,Blas:2010hb} and has been called `khronon', whereas the model itself has been named `khrono-metric' gravity. The jury is out on the renormalizability and ultraviolet behavior of the non-projectable Ho\v rava gravity; some encouraging results have been reported in \cite{Bellorin:2023dwk,Blas:2025ejp}. 

A more general phenomenological model for description of preferred frame effects in gravity was proposed by Jacobson and Mattingly \cite{Jacobson:2000xp} and is known as Einstein--\ae ther (\AE) theory. It is an extension of GR that contains, in addition to the metric, a timelike unit-norm vector field --- \ae ther --- which locally breaks the Lorentz symmetry group down to the group of rotations. The dynamics is described by the most general diff-invariant Lagrangian containing up to two derivatives of the metric and \ae ther \cite{Jacobson:2007veq}. The theory is free from ghost or gradient instabilities and presents a consistent effective field theory at the quantum level \cite{Withers:2009qg}. In addition to the helicity-2 gravitons, it propagates excitations with helicities 0 (scalar) and 1 (vector) around the Minkowski background \cite{Jacobson:2004ts}. In the limit when the \ae ther is constrained to be hypersurface-orthogonal, the vector excitation decouples, and the khrono-metric model is recovered \cite{Jacobson:2013xta}. The supersymmetric 
extension of the \AE-theory was studied in~\cite{Pujolas:2011sk,Marakulin:2016net}.  

The parameter space of the \AE-theory has been strongly constrained through astrophysical and cosmological observations \cite{Oost:2018tcv,Gupta:2021vdj} (see also the discussion in \cite{Adam:2021vsk}) thereby verifying local LI of gravity with high precision. In  particular, an almost simultaneous detection of the GW and $\gamma$-ray signals from a binary neutron star merger \cite{LIGOScientific:2017zic} has established the speed of GWs to be equal to the speed of light within tiny uncertainty, yielding the tightest constraint on a combination of \AE-theory parameters. The constraints on other combinations are less stringent and generally fall short of the most precise tests of LI in particle physics \cite{Mattingly:2005re,Kostelecky:2008ts,Liberati:2013xla}.
It is natural to ask if gravitational waveforms can provide further probes of the \AE-theory and thereby improve tests of LI in gravity.

A complete modeling of waveforms from a binary merger in the \AE-theory would require numerical simulations implementing the dynamics of the \ae ther field into GR codes. While this task appears in principle possible \cite{Sarbach:2019yso}, it is highly challenging. Alternatively, one can focus on the regime when the size of at least one of the companions is much smaller than the orbital radius. Then one can treat the smaller companion as a point particle coupling to the metric and \ae ther through a set of effective worldline operators, along the lines of \cite{Goldberger:2004jt}. This approach can be efficiently combined with the post-Newtonian expansion \cite{Will:2018ont,Taherasghari:2023rwn,Taherasghari:2025mlf} where the interaction of the effective point particle with the \ae ther is described order by order in velocity by the so-called sensitivity parameters \cite{Foster:2007gr}. This approach was adopted in Ref.~\cite{Schumacher:2023cxh} which modeled waveforms from a binary neutron star inspiral in \AE-theory using prior calculations of neutron star sensitivities \cite{Yagi:2013ava,Gupta:2021vdj}. Application to two binary neutron star systems in the LIGO/Virgo data unfortunately did not produce any constraints competitive with already existing bounds. It was emphasized, however, that since most of GW events come from black hole (BH) binaries, the bounds could be improved if calculation of BH sensitivities were available. The BH sensitivities are also needed to exploit the proposal of \cite{Heisenberg:2025tfh} to constrain the \AE-theory using the gravitational memory effect.

In Ref.~\cite{Kovachik:2023fos}, some of us derived the leading-order sensitivity parameter for BHs in the khrono-metric model. In the present paper we extend the calculation to the general \AE-theory. We use the same strategy as in \cite{Kovachik:2023fos}. The sensitivity is extracted from the asymptotics of the solution describing a BH slowly moving with respect to the \ae ther. To find this solution, we exploit the fact that the \AE\ couplings are already constrained to be much less than unity by the existing bounds. This allows us to Taylor expand the metric and the \ae ther field of a BH in these couplings. At the leading order, the back-reaction of the \ae ther on the metric can be neglected and the problem reduces to finding the \ae ther configuration in the standard BH spacetime; this regime is known at the `decoupling limit'. We focus on Schwarzschild BHs; while this is an important simplification compared to the realistic Kerr case, it still may provide a good approximation, given that the majority of LIGO/Virgo primary BHs appear to have relatively low spins \cite{LIGOScientific:2025pvj}. 
Static spherically symmetric solution for a BH at rest with respect to the \ae ther is known and happens to be the same in the \AE\ and khrono-metric theories \cite{Eling:2006ec,Barausse:2011pu,Blas:2011ni}.\footnote{Rotating BHs in 
the \AE-theory were constructed in \cite{Barausse:2015frm} in the limit of low spin; solutions with arbitrary spin were found in \cite{Adam:2021vsk}.}  
We perturb this solution to endow it with a small velocity. The perturbation is linear in velocity and is obtained by numerically solving the linearized \ae ther equations in the original spherically symmetric background. The relevant solution is picked up by imposing 
boundedness at infinity and 
the conditions of regularity at the causal horizons for the scalar and vector modes. Comparing the asymptotics of the BH \ae ther field to that of a point particle, we determine the sensitivity.

We find that the sensitivity is independent of the BH mass, as expected on dimensional grounds. It is proportional to one of the \AE\ parameters, with an order-one proportionality coefficient depending on the speeds of the scalar and vector \ae ther modes. We investigate this dependence numerically and analytically and show that in the limit when the scalar and vector speeds are large, the sensitivity reduces to the value previously found in the khrono-metric model. We derive corrections to this value at finite scalar and vector speeds. The resulting analytic formula provides a very good approximation to the numerically calculated sensitivity throughout all of the presently allowed parameter space. 

The paper is organized as follows. In Sec.~\ref{sec:notationbkg} we review the \AE-theory, summarize the constraints on its parameters, and describe the solution for a Schwarzschild BH at rest with respect to the \ae ther in the decoupling limit. In Sec.~\ref{sec:movingbh} we derive the equations for  a perturbation describing slow motion of the BH relative to the \ae ther and solve them numerically. In Sec.~\ref{sec:sens} we obtain the 
\ae ther field of a moving point particle coupled to the \ae ther and match it to the asymptotics of the moving BH solution; this procedure yields the BH sensitivity. In Sec.~\ref{sec:largespeeds} we analytically study the dependence of the sensitivity on the speeds of the scalar and vector \ae ther modes and obtain a simple analytic approximation, accurate whenever these speeds are greater than unity. We conclude in Sec.~\ref{sec:conc}. Appendices contain technical details of the analysis.

We use the system of units $\hbar=c=1$ and mostly negative metric signature $(+,-,-,-)$.

\section{Einstein--aether model}\label{sec:notationbkg}

\subsection{Lagrangian}
\label{ssec:Lagr}

The \AE-theory \cite{Jacobson:2007veq} introduces a time-like vector $u^\mu$ with unit norm,
\begin{equation}
\label{unitnorm}
    u_\mu u^\mu = 1\;,
\end{equation}
which covariantly couples to the spacetime metric $g_{\mu\nu}$. This vector decsribes a preferred time direction at every point of spacetime and is called \ae ther. The dynamics of the \ae ther is assumed to be governed by an effective field theory admitting a derivative expansion. At the lowest order, \ae ther enters into the Lagrangian through terms with two derivatives. There are in total four nonequivalent terms. Adding the Einstein--Hilbert term for the metric, the action of the \AE-theory reads, 
\begin{equation}
\label{actgen}
    S = \frac{-1}{16\pi G} \int \dd[4]{x} \sqrt{-g} \left(R + \mathcal{L}_{\text{\AE}}\right) ,
\end{equation}
where
\begin{equation}\label{e:AEAction}
    \mc{L}_{\text{\AE}} = 
    c_1 (\nabla_\mu u^\nu)(\nabla^\mu u_\nu)
    +c_2 (\nabla_\mu u^\mu)^2
    +c_3 (\nabla_\mu u^\nu) (\nabla_\nu u^\mu)
    +c_4 (u^\mu \nabla_\mu u^\nu)(u^\lambda \nabla_\lambda u_\nu)\;.
\end{equation}
Here $R$ is the Ricci scalar and $\nabla_\mu$ denotes for the covariant derivative.
The dimensionless constants $c_i$, $i=1,\ldots,4$, describe the strength of the \ae ther interaction with gravity. We will assume them to be small, $c_i\ll 1$, consistently with the phenomenological bounds that will be reviewed shortly. We also assume that the 
\ae ther does not directly couple to the Standard Model matter. The gravitational constant $G$ is related to the Newton constant inferred from the gravitational field of compact weakly gravitating bodies as
\begin{equation}\label{MassGNrelation}
G = G_N\left(1-\frac{c_1+c_4}{2}\right)\;.
\end{equation}
The constants $G$ and $G_N$ essentially coincide in the limit of small \ae ther couplings. The equations of motion of the \AE-theory are obtained by varying the action (\ref{actgen}) with respect to $g_{\mu\nu}$ and $u^\mu$, with the variations subject to the constraint (\ref{unitnorm}).\footnote{Alternatively, the constraint can be implemented by adding a Lagrange multiplier term to the action.} 

A convenient parameterization of the \ae ther Lagrangian elucidating its geometrical content was suggested in Ref.~\cite{Jacobson:2013xta}. One decomposes the gradient of the \ae ther in terms of the acceleration $a_\mu$, expansion $\vartheta$, shear $\sigma_{\mu\nu}$ and twist $\omega_{\mu\nu}$:
\begin{equation}\label{e:gradu}
    \nabla_\mu u_\nu = u_\mu a_\nu+\frac13 \vartheta P_{\mu\nu} + \sigma_{\mu\nu} + \omega_{\mu\nu} ,
\end{equation}
where 
\bseq
\label{alldefs}
\begin{gather}
%\begin{align}
\label{athetPdef}
    a_\mu = u^\nu \nabla_\nu u_\mu\;,\qquad
    \vartheta=\nabla_\mu u^\mu\;,\qquad
    P_{\mu\nu}=g_{\mu\nu}-u_\mu u_\nu\;,\\
    \label{sigomdef}
    \sigma_{\mu\nu}=\nabla_{(\mu}u_{\nu)}-u_{(\mu}a_{\nu)}-\frac{\vartheta}{3}P_{\mu\nu}\;,\qquad
    \omega_{\mu\nu}=\nabla_{[\mu} u_{\nu]}-u_{[\mu}a_{\nu]}\;,
%\end{align}
\end{gather}
\eseq
and round (square) brackets denote symmetrization (anti-symmetrization) over indices. The unit-norm condition (\ref{unitnorm}) implies the identities,
\be
\label{idents}
u^\mu a_\mu=u^\mu P_{\mu\nu}=u^\mu \sigma_{\mu\nu}=u^\mu \omega_{\mu\nu}=0\;.
\ee
Substituting (\ref{e:gradu}) into (\ref{e:AEAction}) and using these identities, one obtains
\be
\label{e:AEAction1}
{\cal L}_{\text{\AE}}=\frac{\tilde c_\vartheta}{3}\,\vartheta^2+c_\sigma\,\sigma_{\mu\nu}\sigma^{\mu\nu}
+\tilde c_\omega\,\omega_{\mu\nu}\omega^{\mu\nu}+c_a \,a_\mu a^\mu
\ee
with the new couplings
\be
\label{ccs}
\tilde c_\vartheta=c_1+3c_2+c_3\;,\qquad c_\sigma =c_1+c_3\;,\qquad \tilde c_\omega=c_1-c_3\;,\qquad c_a=c_1+c_4\;.
\ee

Yet another equivalent form of the Lagrangian can be obtained as follows. 
We integrate by parts the third term in the 
original Lagrangian (\ref{e:AEAction}), giving 
\be
\label{byparts}
c_3 (\nabla_\mu u^\nu) (\nabla_\nu u^\mu)\simeq
-c_1(\nabla_\mu u^\nu) (\nabla_\nu u^\mu)+(c_1+c_3)
(\nabla_\mu u^\mu)^2-(c_1+c_3) R_{\mu\nu} u^\mu u^\nu\;.
\ee
Then the substitution (\ref{e:gradu}) leads to
\be
\label{e:AEAction2}
{\cal L}_{\text{\AE}}=\frac{c_\vartheta}{3}\,\vartheta^2
+ c_\omega\,\omega_{\mu\nu}\omega^{\mu\nu}+c_a \,a_\mu a^\mu
-c_\sigma R_{\mu \nu}u^\mu u^\nu
\ee
with 
\be
\label{ccsnew}
c_\vartheta = \tilde c_\vartheta+2c_\sigma\;,~~~~
 c_\omega=\tilde c_\omega+c_\sigma\;.
\ee
Elimination of the $\sigma_{\mu\nu}$-term significantly simplifies the calculations. Further, the last term in (\ref{e:AEAction2}) vanishes in Ricci-flat backgrounds, such as Minkowski, Schwarzschild or Kerr metrics. Thus, the dynamics of the \ae ther in such backgrounds is described by only three couplings $c_\theta$, $c_\omega$ and $c_a$.
In what follows we work with the form (\ref{e:AEAction2}) of the \ae ther Lagrangian.

In the Minkowski spacetime with constant background value of the \ae ther, $\bar u^\mu=(1,0,0,0)$, the \AE-theory propagates 
two tensor (helicity-2) graviton polarizations with the squared propagation speed  
\cite{Jacobson:2004ts}
\be
\label{Tspeed}
s_T^2=\frac{1}{1-c_\sigma}\;.
\ee
In addition, there are two vector (helicity-1) and one scalar (helicity-0) modes. The leading-order expressions for their squared propagation speeds
in the limit of small \ae ther couplings are
\be
\label{speeddefns}
s_V^2=\frac{c_\omega}{2c_a}\;,\qquad
s_S^2=\frac{c_\vartheta}{3c_a}\;.
\ee
Note that in general all speeds are different from unity (the speed of light), which is a consequence of LI violation by the background \ae ther field.

It is instructive to momentarily relax our assumption that the parameters $c_i$ are small and consider the limit \cite{Jacobson:2013xta}
\be
\label{khronolimit}
c_\omega \to \infty\;.
\ee
In this limit, the twist $\omega_{\mu\nu}$ is forced to vanish, implying that the \ae ther must be orthogonal to a family of spacelike hypersurfaces. Parameterizing the latter as the levels of a scalar function $\vf(x)$, we can write
\be
\label{khronoconnect}
u^\mu=\frac{\nabla^\mu\vf}{\sqrt{\nabla_\nu\vf\nabla^\nu\vf}}\;.
\ee
Note that the vector \ae ther polarization has disappeared. Formally, the propagation speed of vector excitations $s_V$ becomes infinite in the limit (\ref{khronolimit}). We can take $\vf(x)$ as the new dynamical field. It defines a preferred time variable and has been called khronon. Substituting the relation (\ref{khronoconnect}) back into the Lagrangian (\ref{e:AEAction2}) and supplementing it with the Einstein--Hilbert term, we obtain the Lagrangian of khrono-metric gravity \cite{Blas:2010hb} with the coupling identification (in the notations of \cite{Blas:2010hb})
\be
\label{kronocouplings}
c_\vartheta \leftrightarrow 3(\l'+\b)\;,\qquad 
c_a \leftrightarrow \a\;,\qquad
c_\sigma \leftrightarrow \b\;.
\ee
Khrono-metric gravity represents the low-energy limit of the power-counting renormalizable non-projectable Ho\v rava gravity \cite{Horava:2009uw,Blas:2009qj}.

\begin{figure}[t]
    \centering
~\hspace{0.5cm}
\includegraphics[width=0.9\linewidth]{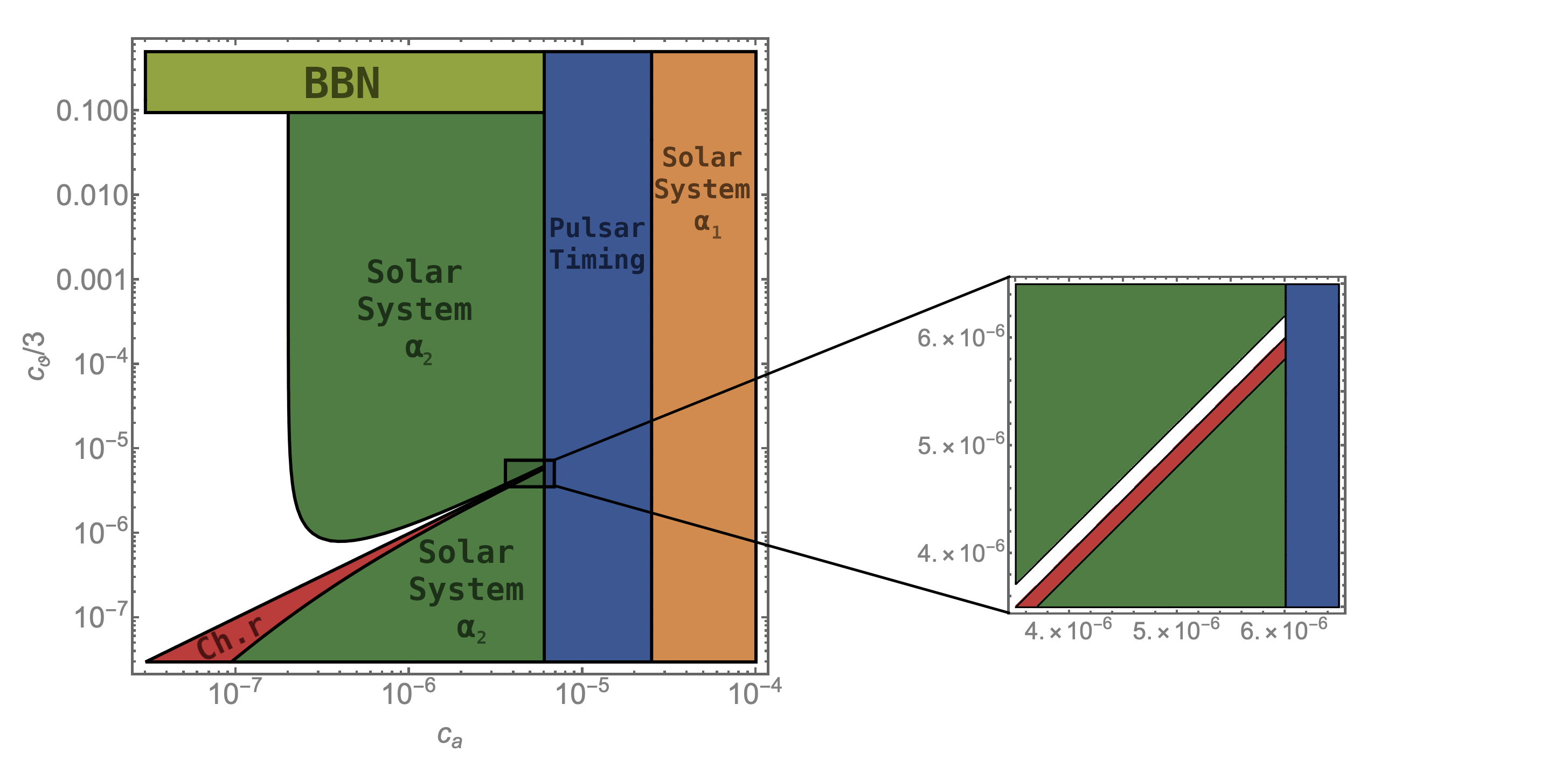}
    \caption{Allowed parameter space for the \ae ther couplings $c_a$, $c_\vartheta$  (white). The colored regions are excluded by 
     the primordial helium abundance (light green), constraints on the post-Newtonian parameters from the Solar System (dark green and orange), timing of the triple pulsar system (blue),  absence of Cherenkov radiation losses by cosmic rays (red). The inset shows a zoom-in on the region around the line $c_\vartheta/3=c_a$ corresponding to unit propagation velocity of scalar \ae ther mode, $s_S=1$.
}
    \label{fig:constr}
\end{figure}

The parameters of the \ae ther Lagrangian are subject to theoretical and observational constraints \cite{Oost:2018tcv}. Positivity of energy and stability of the \ae ther excitations require $c_\vartheta$, $c_\omega$ and $c_a$ to be positive. The measurement of the speed of GWs from the binary neutron star merger event GW170817 / GRB 170817A \cite{LIGOScientific:2017zic} puts a very stringent bound on the parameter~$c_\sigma$,
\be
\label{constrGW}
|c_\sigma|\lesssim 10^{-15}\;.
\ee
Further, a subluminal propagation of any of the \ae ther modes would lead to their strong Cherenkov-type emission by ultra-high-energy cosmic rays \cite{Elliott:2005va}. Absence of the related energy losses implies one-sided bounds,
\be
\label{constrCheren}
s_{S,V}\geq 1~~~~ \Longleftrightarrow~~~~ c_\vartheta\geq 3c_a\;,~~~c_\omega\geq 2 c_a\;.
\ee
The primordial helium abundance tied to the expansion rate of the universe during the big bang nucleosynthesis (BBN) gives an upper bound on $c_\vartheta$ \cite{Carroll:2004ai}
\be
\label{constrBBN}
c_\vartheta \lesssim 0.3\;.
\ee
The rest of the constraints come from
weak-field gravitational effects of the \AE-theory. These are conveniently analyzed within the parameterized post-Newtonian (PPN) formalism \cite{Foster:2005dk}. Within this framework, the \AE-theory differs from GR by only two parameters characterizing LI violation, which under the assumptions $c_\sigma\ll c_a\ll 1$ have the form,\footnote{Exact expressions can be found in \cite{Foster:2005dk}.} 
\begin{align}\label{alpha12Def}
\alpha^\text{PPN}_1 = -4c_a\;,\qquad
\alpha^\text{PPN}_2  =-\frac{c_a}{2}\left( 1-\frac{3c_a}{c_\vartheta}\right)\;.
\end{align}
Note that the second of these parameters vanishes if the speed of the scalar \ae ther polarization coincides with the speed of light, $s_S=1$. Bounds on these parameters from Solar System observations are $|\alpha_1^\text{PPN}|\lesssim 10^{-4}$, 
$|\alpha_2^\text{PPN}|\lesssim 10^{-7}$ \cite{Will:2005va}. Timing of a triple pulsar system PSR~J0337+1715 allows one to push the first of these bounds down to 
 $\abs{\alpha_1^\text{PPN}}< 2.4\times 10^{-5}$ at $95\%$ CL \cite{Gupta:2021vdj}.\footnote{A recent robust timing analysis of a different system PSR J1738+0333 yields somewhat weaker bound $\abs{\alpha_1^\text{PPN}}< 7.2\times 10^{-5}$ at $90\%$ CL \cite{Vaglio:2026vmo}. 
 }
 The parameter space of the couplings $c_a$, $c_\vartheta$ allowed by these and other constraints is shown in Fig.~\ref{fig:constr}. The remaining parameter $c_\omega$ is unconstrained, apart from the lower bound in (\ref{constrCheren}).

We have seen that the couplings of the \AE-theory, except perhaps $c_\omega$, must be small. This simplifies the analysis of the \AE\ equations. The energy-momentum tensor of the \ae ther obtained from the Lagrangian (\ref{e:AEAction2}) is proportional to the \ae ther couplings,
$T^{\mu\nu}_\text{\AE}={\cal O}(c_i)$.
If $c_i$ are small,\footnote{For the sake of the argument, we assume here that $c_\omega$ is small, as the rest of the couplings. However, this is not needed: the khrono-metric theory obtained in the limit (\ref{khronolimit}) also admits the perturbative analysis as long as $c_{\vartheta,a,\sigma}\ll 1$ \cite{Kovachik:2023fos}.} 
one can neglect $T^{\mu\nu}_\text{\AE}$ in the equations for the metric, which thus takes the same form as in GR. The \ae ther equations are then solved in this fixed geometry. In other words, the metric and \ae ther equations decouple, so this approximation is called decoupling limit. At the next step of the perturbative procedure, one can compute the corrections to the metric by solving the Einstein's equations with $T^{\mu\nu}_\text{\AE}$ evaluated on the decoupling-limit solution. This step, however, is often unnecessary: we will see below that the decoupling-limit solution is sufficient to extract physically relevant information.

\subsection{Static black holes in the decoupling limit}
\label{sec:stationary}

Spherically symmetric BHs in the \AE-theory were obtained in \cite{Eling:2006ec,Barausse:2011pu} by solving the full set of metric and \ae ther equations. Ref.~\cite{Blas:2011ni} derived spherical BHs in the decoupling limit of the khrono-metric model and showed that they are the same as BHs in the decoupling limit of the general \AE-theory. These solutions describe Schwarzschild BHs at rest with respect to the \ae ther reference frame and it has been verified that they form as a result of spherical gravitational collapse \cite{Garfinkle:2007bk,Saravani:2013kva,Bhattacharyya:2015uxt,Bhattacharjee:2018nus,Franchini:2021bpt}. Let us briefly review their properties.

Neglecting the effect of the \ae ther on the metric, the latter takes the standard Schwarz\-schild 
form,
\be
\label{metrr}
\dd s^2=\tilde f(r)\,\dd t^2 -\frac{\dd r^2}{\tilde f(r)}-r^2 d\Omega^2\;, \qquad
\tilde f(r)=1-\frac{2Gm}{r}\;,
\ee
where $\dd \Omega^2=\dd\theta^2+\sin^2\theta \,\dd\phi^2$ is the line element on a unit sphere and $m$ is the BH mass. 
In the analytic expressions below we will not use the concrete form of the metric function $\tilde f(r)$, specifying it only in the numerical calculations.
 This makes the formulas less cluttered and admits immediate generalization of our results to other spherically symmetric metrics. 
We set the units of length such that 
\be
\label{lengthunits}
2G m=1
\ee
and introduce the coordinate $\xi=1/r$ which brings the metric into the form,
\be
\label{metrxi}
\dd s^2=f(\xi)\,\dd t^2 -\frac{\dd \xi^2}{\xi^4f(\xi)}-\frac{ d\Omega^2}{\xi^2}\;,\qquad f(\xi)\equiv\tilde f(1/\xi)\;.
\ee
Note that in the new coordinates, the asymptotic infinity is located at $\xi=0$, and the Schwarzschild horizon is at $\xi=1$.

Under the condition of spherical symmetry,
the \ae ther has only the temporal and radial components which we parameterize as
\be
\label{ubar}
\bar u_t=U\;,\qquad\bar u^r=V~~~\Longleftrightarrow~~~\bar u^\xi=-\xi^2V\;.
\ee
Here and in what follows we use an overbar to highlight spherically symmetric configurations. Note that for a BH the \ae ther field if tilted towards the center, so $V$ is negative, $V<0$. At the asymptotic infinity, the \ae ther becomes purely time-directed,
\be
\label{UVasympt}
U\to 1\;,~~~~V\to 0 ~~~~ \text{at}~~\xi\to 0\;.
\ee
In terms of the functions $U(\xi)$, $V(\xi)$ the unit-norm constraint reads,
\be
\label{UVunit}
U^2-V^2=f\;,
\ee
implying that only one of the \ae ther components is independent.

\begin{figure}[t]
    \centering
%    \vspace{-0.5cm}
    \includegraphics[width=0.5\linewidth]{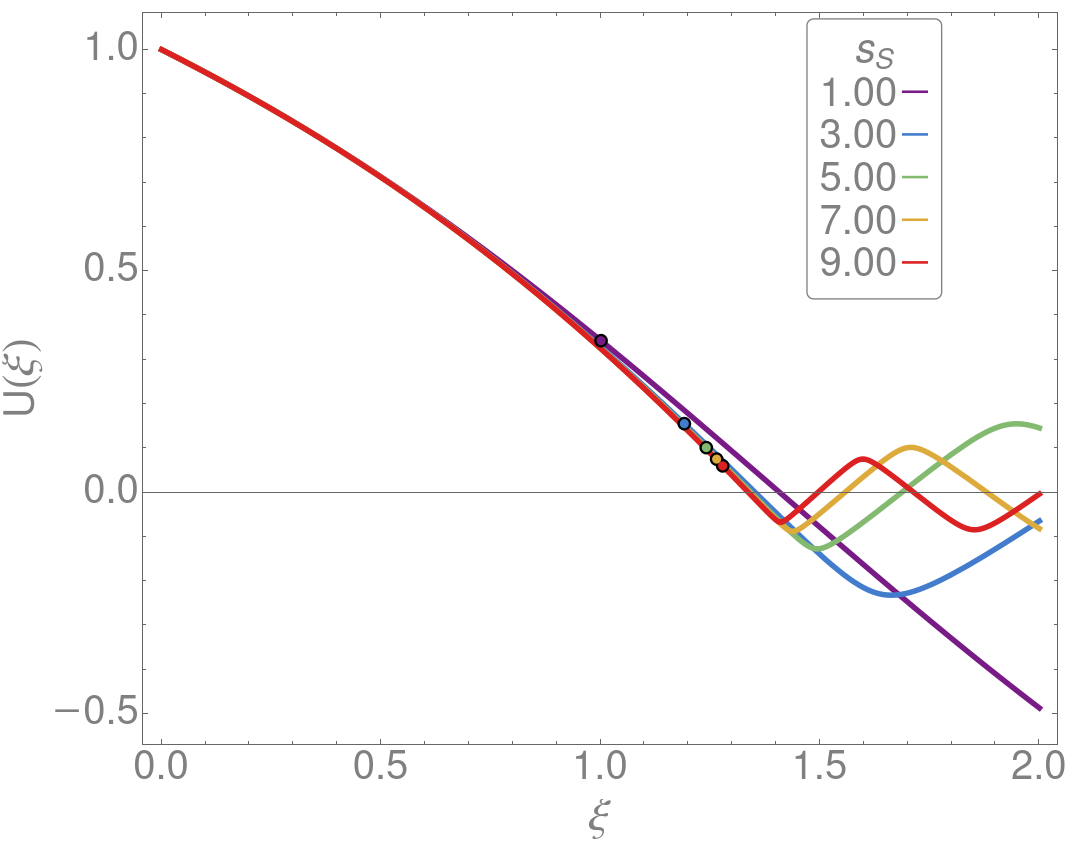}
    \caption{Temporal component of the \ae ther for a static Schwarzchild black hole, shown as function of the variable $\xi=2Gm/r$ for several values of the helicity-0 speed $s_S$. Colored dots mark the position of the helicity-0 horizon for each curve. The curves do not depend on the helicity-1 speed $s_V$.}
    \label{fig:backgroundsolution}
\end{figure}

The values of the expansion, acceleration and twist are found to be 
\be
\label{abck}
\bar \vartheta =-\xi^2 V'+2\xi V\;,~~~~
\bar a_t=-\xi^2VU'~,~~~~\bar a_\xi=-\frac{UU'}{f}~,~~~~
\bar a_\theta=\bar a_\phi=0\;,~~~~
\bar\omega_{\mu\nu}=0\;,
\ee
where prime stands for derivative with respect to $\xi$. Substitution of these expressions into the \ae ther action and integration by parts in $\xi$ yields, 
\be
\bar S_\text{\AE}=-\frac{1}{4G}\int \dd t\dd\xi \bigg[\frac{c_\vartheta}{3}\bigg(V'^2+\frac{2V^2}{\xi^2}\bigg)-c_a U'^2\bigg]\;.
\ee
Equation satisfied by the \ae ther is obtained by varying this action with respect to $U$ and $V$ subject to the constraint (\ref{UVunit}). It reads, 
\bseq
\label{Ueq}
\be
\label{Ueq1}
\frac{U''}{U}-s^2_S\bigg(\frac{V''}{V}-\frac{2}{\xi^2}\bigg)=0\;,
\ee
or substituting $V$ from (\ref{UVunit}),
\be
\label{Ueq2}
U''+\frac{s_S^2 U}{(1-s_S^2)U^2-f}\bigg[-U'^2
+\frac{(UU'-f'/2)^2}{U^2-f}+\frac{2(U^2-f)}{\xi^2}+\frac{f''}{2}\bigg]=0\;.
\ee
\eseq
Note that this equation depends only on the speed of the scalar \ae ther perturbations $s_S$. This is the same equation as in the khrono-metric model studied in 
\cite{Blas:2011ni,Kovachik:2023fos}. The denominator of the second term vanishes at $\xi_S$ such that 
\be
\label{Shoriz}
(s_S^2-1)U^2(\xi_S)+f(\xi_S)=0\;.
\ee
This corresponds to the causal horizon for the helicity-0 mode. Indeed, the 
scalar \ae ther excitations propagate in the effective metric \cite{Eling:2006ec}
\be
\label{Smetr}
g_{\mu\nu}^{(S)}=g_{\mu\nu}+(s_S^2-1)\bar u_\mu \bar u_\nu\;.
\ee
Clearly, Eq.~(\ref{Shoriz}) coincides with the condition $g_{tt}^{(S)}=0$, implying that the surface $\xi=\xi_S$ is lightlike in this metric. The condition that the solution must be regular at $\xi_S$, together with the boundary condition (\ref{UVasympt}), fixes the solution uniquely. 
Equation (\ref{Ueq}) cannot be solved analytically. The results of numerical integration for several values of $s_S$ are shown in Fig.~\ref{fig:backgroundsolution}. These solutions are regular everywhere at $\xi>0$ and coincide with those found in the previous studies \cite{Blas:2011ni,Kovachik:2023fos}.
Below we will need the derivative of the function 
$U(\xi)$ at $\xi\to 0$. This can be read off directly from Eq.~(\ref{Ueq2}) by using the regularity of $U''(0)$. We obtain 
\be
\label{Uder0}
U'(0)=f'(0)/2\;,
\ee
independently from the value of $s_S$. For the Schwarzchild metric, $f(\xi)=1-\xi$, it becomes $U'(0)=-1/2$.

Although the solutions do not depend on the helicity-1 propagation speed $s_V$, they possess a horizon for this mode. Similarly to (\ref{Smetr}), the latter mode propagates in the effective metric
\be
\label{Vmetr}
g_{\mu\nu}^{(V)}=g_{\mu\nu}+(s_V^2-1)\bar u_\mu \bar u_\nu\;,
\ee
and thus its horizon is located at $\xi_V$ defined by the equation
\be
\label{Vhoriz}
(s_V^2-1)U^2(\xi_V)+f(\xi_V)=0\;.
\ee
If $s_V$ is smaller (bigger) than $s_S$, we have $\xi_V<\xi_S$ ($\xi_V>\xi_S$). In terms of the radial coordinate $r$, this corresponds to the helicity-1 horizon lying outside (inside) the helicity-0 horizon.

If $s_V$ is large, there are additional helicity-1 horizons appearing at large values of $\xi$ where the function $U(\xi)$ oscillates, see Fig.~\ref{fig:backgroundsolution}. These inner horizons are, however, cloaked by the outer scalar, vector and Schwarzschild horizons and thus are causally disconnected from infinity. 
Throughout this paper we denote by $\xi_V$ the smallest root of Eq.~(\ref{Vhoriz}) which corresponds to the outermost vector horizon.

Note that both scalar and vector horizons lie at $\xi$ less than $\xi_\star$, where $\xi_\star$ is the smallest root of the solution $U(\xi)$. In fact, no matter how fast an excitation moves, it cannot escape to infinity from the interior of the surface $r=1/\xi_\star$ which thus forms a {\it universal horizon} \cite{Barausse:2011pu,Blas:2011ni}. One can generalize the notion of universal horizon to static asymptotically flat geometries which are not necessarily spherically symmetric \cite{Bhattacharyya:2015gwa}: it is a compact hypersurface orthogonal to the \ae ther and tangential to the Killing vector of time translations. 
While the universal horizon is regular for spherical BHs, thermodynamic arguments \cite{Dubovsky:2006vk} suggest that it may be unstable in the khrono-metric theory with respect to non-linear aspherical perturbations \cite{Blas:2011ni}.
 Non-analyticity of the khronon at the universal horizon of the slowly moving BHs found in \cite{Kovachik:2023fos} may be an indication of such instability. Further, the studies of rotating BHs in the \AE-theory at finite values of the excitation speeds $s_S$, $s_V$ indicate that they do not possess a universal horizon \cite{Barausse:2015frm,Adam:2021vsk}. Below we address existence of the universal horizon for slowly moving BHs in the \AE-theory.

\section{Slowly moving black holes}\label{sec:movingbh}

\subsection{\AE ther in the black hole rest frame}
We now consider a BH that moves with respect to the \ae ther with velocity ${\bf v}$ which, without loss of generality, we assume to be directed along the $z$-axis,
\be
\label{velocity}
v^x=v^y=0\;,~~~~v^z=v\;.
\ee
We also assume that the motion is slow, $v\ll 1$. Since LI is broken, we cannot simply change coordinates in the static solution to arrive at the moving BH. However, we can still choose the coordinate frame co-moving with the BH. The solution in this frame is time independent and the metric remains Schwarzschild.
This comes at the expense of modifying the asymptotic value of the \ae ther field. For small velocity, the modification is small, so we look for the new \ae ther embedding in the form,
\be
\label{aetherpert}
u^\mu=\bar u^\mu +b^\mu\;,
\ee
where $\bar u^\mu$ is the spherically symmetric solution from Sec.~\ref{sec:stationary} and $b^\mu$ is a small perturbation. At large distance from the BH we have,
\be
\label{basymp}
b^x=b^y=0\;,~~~b^z=-v~~~~\Longleftrightarrow~~~~b^r=-v \cos\theta\;,~~~
b^\theta=v\frac{\sin\theta}{r}\;,~~~b^\phi=0\;~~~~\text{at}~~r\to\infty\;.
\ee
This suggests taking $b^\mu$ in the form of a dipole perturbation,
\begin{align}
b_\mu \dd{x}^\mu &=
b_t \dd{t}
+ v \frac{U}{f} B_1 \cos\theta \dd{r}
- v r B_2 \sin\theta \dd{\theta}\notag\\
&=b_t \dd{t}
- v \frac{U}{\xi^2 f}B_1 \cos\theta \dd{\xi}
- v \frac{B_2}{\xi} \sin\theta \dd{\theta}\;,
\label{bdipole}
\end{align}
where the radial functions $B_1(\xi)$, $B_2(\xi)$ satisfy the boundary conditions
\be
\label{BBbc}
B_1(\xi)\,,~B_2(\xi)~\to~ 1~~~\text{at}~~\xi\to 0\;.
\ee
Note that for the later convenience, we have inserted a factor
$U/f$ in the definition of the radial component of the perturbation. 
One might be worried that division by the metric function $f$ can introduce spurious singularity at the Schwarzschild horizon. This does not happen. Indeed, raising the radial index, we obtain  
\be
\label{bhatB}
b^r=-vU B_1\cos\theta\;.
\ee
The contravariant radial component of any vector is the same in all coordinate frames that share the same radial coordinate. In particular, it is the same in the Schwarzschild and the Finkelstein frames which differ only by the choice of the time variable. 
Since the Finkelstein frame is regular at the Schwarzschild horizon, $b^r$ must be also regular there. Further, from Fig.~\ref{fig:backgroundsolution} we see that the background function $U(\xi)$ is non-zero at $\xi=1$. Hence, $B_1$ is analytic at this point.

Due to the spherical symmetry of the background, perturbations from different multipole sectors decouple, so the Ansatz (\ref{bdipole}) is compatible with the linearized \ae ther equations. 
The temporal component $b_t$ is not independent: it is determined from the unit-norm constraint~(\ref{unitnorm}).
Solving this constraint with respect to $b_t$, we obtain,
\be
\label{btexp}
b_t=-vV B_1\cos\theta
+v^2\frac{f}{2U}B_1^2\cos^2\theta
+v^2\frac{f}{2U} B_2^2\sin^2\theta+{\cal O}(v^3)\;,
\ee
where we have retained terms up to quadratic order in $v$.

We proceed to deriving equations satisfied by the functions $B_1(\xi)$, $B_2(\xi)$. To this aim, we compute the 
quadratic action for the \ae ther perturbation (\ref{bdipole}).

\subsection{Quadratic action for \ae ther perturbation}

The quadratic action has order ${\cal O}(v^2)$, hence we need to expand various quantities appearing in the \ae ther Lagrangian (\ref{e:AEAction2}) up to this order. Note that the last term proportional to $c_\sigma$ vanishes because the Schwarzschild metric is Ricci flat. We compute the three remaining terms one by one.

\subsubsection{$\vartheta$-term}

The expansion $\vartheta$ contains only linear perturbation,
\be
\label{varthetalin}
\vartheta^{(1)}=v\cos\theta \big[\xi^2 U B'_1+(\xi^2 U'-2\xi U)B_1+2\xi B_2 \big]\;,\qquad \vartheta^{(2)}=0\;.
\ee
Squaring this expression, integrating over angles, and simplifying with integration by parts in the variable $\xi$, we get the $\vartheta$ term in the action,
\be
\label{Svartheta}
S_{\vartheta}^{(2)}=-\frac{v^2}{12G}
\int \dd t\dd\xi\,\frac{c_\vartheta}{3}\bigg[
U^2 B_1'^2-\frac{4U}{\xi}B_1B_2'
+\bigg(\frac{2U^2}{\xi^2}-U U''\bigg) B_1^2
-\frac{4U}{\xi^2}B_1B_2+\frac{4}{\xi^2}B_2^2
\bigg].
\ee

\subsubsection{$\omega$-term}

Since the background value of the twist $\omega_{\mu\nu}$ vanishes, 
there are no terms in the action involving its second variation that would have the form $\bar \omega^{\mu\nu}\omega^{(2)}_{\mu\nu}$. Thus it suffices to compute the linear perturbation $\omega^{(1)}_{\mu\nu}$: 
squaring it will give the quadratic action. 
Using the definition (second equation in (\ref{sigomdef})) we find,
\be
\label{omegalin}
\omega_{t\theta}^{(1)}=\frac{v}{2}\sin\theta\, V\, \hat\omega\;,\qquad
\omega_{\xi\theta}^{(1)}=\frac{v}{2}\sin\theta \,\frac{U}{\xi^2 f}\, \hat\omega\;,\qquad
\hat\omega\equiv-\xi U B_2'-B_1 +(U+\xi U')B_2\;.
\ee
All other components vanish.
Squaring, integrating over angles and integrating by parts in $\xi$ we obtain,
\be
\label{Somega}
\begin{split}
S_\omega^{(2)}=-\frac{v^2}{12G}\int \dd t\dd\xi\,c_\omega
\bigg[&U^2B_2'^2+\frac{2U}{\xi}B_1B_2' 
+\frac{B_1^2}{\xi^2}-\bigg(\frac{2U}{\xi^2}+\frac{2U'}{\xi }\bigg)B_1B_2\\
&+\bigg(UU''+2U'^2+\frac{4UU'}{\xi}\bigg)B_2^2
\bigg]\;.
\end{split}
\ee

\subsubsection{$a$-term}

Computation of this term is greatly simplified by use of the identity
\be
\label{au}
a_\mu a^\mu=\frac{1}{2} u_{\mu\nu}u^{\mu\nu}-2\omega_{\mu\nu}\omega^{\mu\nu}~,\qquad u_{\mu\nu}=2\nabla_{[\mu} u_{\nu]}\;,
\ee
which is a consequence of the unit-norm constraint (\ref{unitnorm}). 
Unlike $a_\mu$, the anti-symmetric `field strength' $u_{\mu\nu}$ is a linear function of the \ae ther.
We have already computed the $(\omega_{\mu\nu})^2$ contribution, so it remains to find $(u_{\mu\nu})^2$. Only the component $u_{t\xi}$ has non-vanishing value in the background, $\bar u_{t\xi}=-U'$. This means that only this component needs to be expanded up to the second order, whereas for all other components the linear order suffices. A straightforward calculation yields the non-zero contributions:
\bseq
\label{alluperts]}
\begin{align}
&u_{t\xi}^{(1)}=v\cos\theta (V B_1'+V'B_1)\;,\\
&u_{t\xi}^{(2)}=-v^2\bigg[\cos^2\theta\bigg(\frac{f}{2U}B_1^2\bigg)'
+\sin^2\theta\bigg(\frac{f}{2U}B_2^2\bigg)'\,\bigg]\;,\\
&u_{t\theta}^{(1)}=-v\sin\theta\, VB_1\;,\\
&u_{\xi\theta}^{(1)}=v\sin\theta\bigg[-\frac{1}{\xi}B_2'-\frac{U}{\xi^2 f}B_1
+\frac{1}{\xi^2}B_2\bigg]\;.
\end{align}
\eseq
Inserting this into the action and simplifying with the use of integration by parts and Eq.~(\ref{UVunit}), we obtain
\be
\label{Sa}
\begin{split}
S_a^{(2)}=-\frac{v^2}{12G}\int\dd t\dd\xi\,c_a\bigg[&-V^2B_1'^2
-2V^2 B_2'^2
+\bigg(VV''+\frac{fU''}{U}
\bigg)B_1^2
+\frac{4U'}{\xi}B_1B_2\\
&+\bigg(-\frac{2V^2U''}{U}+\frac{2f'}{\xi}-4U'^2-\frac{8UU'}{\xi}
\bigg)B_2^2
\bigg]\;.
\end{split}
\ee

\paragraph{Total.}
Summing up the above terms we arrive at our final expression for the total quadratic \ae ther action,
\be
\label{Sperttot}
\begin{split}
S^{(2)}_\text{\AE}=&-\frac{v^2}{12G}\int\dd t\dd\xi\bigg\{\bigg[c_\vartheta\frac{U^2}{3}-c_aV^2\bigg]B_1'^2+\big[c_\omega U^2-2c_a V^2\big]B_2'^2
+\bigg[-\frac{2c_\vartheta}{3}+c_\omega\bigg]\frac{2U}{\xi}B_1B_2'\\
&+\bigg[c_\vartheta\bigg(-\frac{UU''}{3}+\frac{2U^2}{3\xi^2}\bigg)
+\frac{c_\omega}{\xi^2}+c_a\bigg(VV''+\frac{fU''}{U}\bigg)\bigg] B_1^2\\
&+\bigg[-c_\vartheta\frac{4U}{3\xi^2}
-c_\omega\bigg(\frac{2U'}{\xi}+\frac{2U}{\xi^2}\bigg)
+c_a\frac{4U'}{\xi}\bigg] B_1B_2\\
&+\bigg[c_\vartheta\frac{4}{3\xi^2}+c_\omega\bigg(UU''+2U'^2+\frac{4UU'}{\xi}\bigg)+c_a\bigg(-\frac{2V^2U''}{U}+\frac{2f'}{\xi}-4U'^2-\frac{8UU'}{\xi}\bigg)\bigg]B_2^2\bigg\}.
\end{split}
\ee

\subsection{Equations and solutions}
\label{sec:singular points}

Taking variation of the action (\ref{Sperttot}) with respect to the functions $B_1$ and $B_2$, we obtain a set of two coupled linear equations,
\bseq
\label{perteqs}
\begin{align}
-&\big[(s_S^2 U^2- V^2)B_1'\big]'-2(s_S^2-s_V^2)\frac{U}{\xi}B_2'
+\bigg[s_S^2\bigg(\frac{2U^2}{\xi^2}-UU''\bigg)
+\frac{2 s_V^2}{\xi^2}+VV''+\frac{fU''}{U}\bigg] B_1\notag\\
&-2\bigg[s_S^2\frac{U}{\xi^2}
+s_V^2\bigg(\frac{U'}{\xi}+\frac{U}{\xi^2}\bigg)-\frac{U'}{\xi}
\bigg]B_2=0\;,
\label{perteqs1}\\
\nonumber
\\
-&\big[(s_V^2 U^2-V^2)B_2'\big]'
+(s_S^2-s_V^2)\frac{U}{\xi} B_1'
+\bigg[s_S^2\bigg(\frac{U'}{\xi}-\frac{2U}{\xi^2}\bigg)
-s_V^2\frac{2U'}{\xi}
+\frac{U'}{\xi}\bigg] B_1\notag\\
&+\bigg[s_S^2\frac{2}{\xi^2}+s_V^2\bigg(UU''+2U'^2+\frac{4UU'}{\xi}\bigg)-\frac{V^2U''}{U}+\frac{f'}{\xi}-2U'^2-\frac{4UU'}{\xi}\bigg]B_2=0\;.
\label{perteqs2}
\end{align}
\eseq
Here we have factored out the coefficient $c_a$, which
has allowed us to express all terms through just two parameters: the propagation speeds of the scalar and vector modes (\ref{speeddefns}). Recall that the background functions $U$ and $V$ also implicitly depend on $s_S$, but not on $s_V$.

We solve the system (\ref{perteqs}) numerically. To do this, we first need to identify its singular points. One such point is $\xi=0$ where the coefficient functions in the equations explicitly diverge. This point corresponds to the asymptotic infinity.
Other singular points occur whenever the coefficient in front of the highest derivative in either of the equations vanish, $s_S^2 U^2-V^2=0$ or 
$s_V^2 U^2-V^2=0$. There are two such relevant points lying outside of the universal horizon of the unperturbed BH and corresponding to its scalar and vector horizons, $\xi_S$ and $\xi_V$, respectively; see Eqs.~(\ref{Shoriz}), (\ref{Vhoriz}).
Frobenius-type analysis presented in Appendix~\ref{sec:logtermneed} reveals that at each singular point, one of the four linearly independent solutions of Eqs.~(\ref{perteqs}) diverges.\footnote{We assume here the generic case $\xi_S\neq \xi_V$. If the two horizons coincide, two linearly independent solutions diverge at $\xi=\xi_S=\xi_V$. Eliminating them leads to two boundary conditions.} 
Requiring that the global solution is regular at $\{0,\xi_S,\xi_V\}$ thus gives three conditions which reduce the number of free parameters from four in the general solution of (\ref{perteqs}) down to one. The remaining parameter corresponds to the overall normalization and is fixed by the boundary conditions (\ref{BBbc}).
 This counting shows that for every pair of the excitation speed $(s_S,s_V)$, there is a unique solution of the system (\ref{perteqs}) regular at the asymptotic infinity and the causal horizons $\xi_S$, $\xi_V$.
 Note that the solution may still diverge deep inside the BH if there is an additional inner  
helicity-1 horizon mentioned in Sec.~\ref{sec:stationary}. This singularity is, however, irrelevant for the physics outside the BH since it is cloaked by the outermost tensor (Schwarzschild), vector and scalar horizons; cf.~\cite{Barausse:2015frm}.

On the other hand the system (\ref{perteqs}) does not have any singularity at the universal horizon of the unperturbed solution $\xi_\star$. Indeed, all background functions $U$, $V$ and $f$ are regular at $\xi_\star$, and both highest-derivative coefficients are $-V^2(\xi_\star)=f(\xi_\star)\neq 0$. The only potentially dangerous terms are those with the function $U$ in the denominator, since $U(\xi_\star)=0$. However, all such terms in (\ref{perteqs}) contain the ratio $U''/U$ which is regular at $\xi_\star$, as implied by Eq.~(\ref{Ueq}). We thus conclude that the \ae ther perturbation $B_1$, $B_2$ is regular at the universal horizon. This is in stark contrast with the case of the khrono-metric model where the perturbation possesses a branch-cut singularity at the universal horizon and cannot be extended to $\xi>\xi_\star$ \cite{Kovachik:2023fos}.

To construct the solutions explicitly, we use a version of the shooting method, similar to the one employed in \cite{Kovachik:2023fos}. The solutions are integrated numerically starting from the singular points $\{0,\xi_S,\xi_V\}$ with the conditions of regularity at these points imposed. The resulting solutions are then matched in the intervals between the singular points. The details of the procedure are described in Appendix~\ref{allcoefficients}. 

\begin{figure}[t]
  \centering
  % row 1
  \hspace{-0.5cm}
    \includegraphics[width=0.5\linewidth]{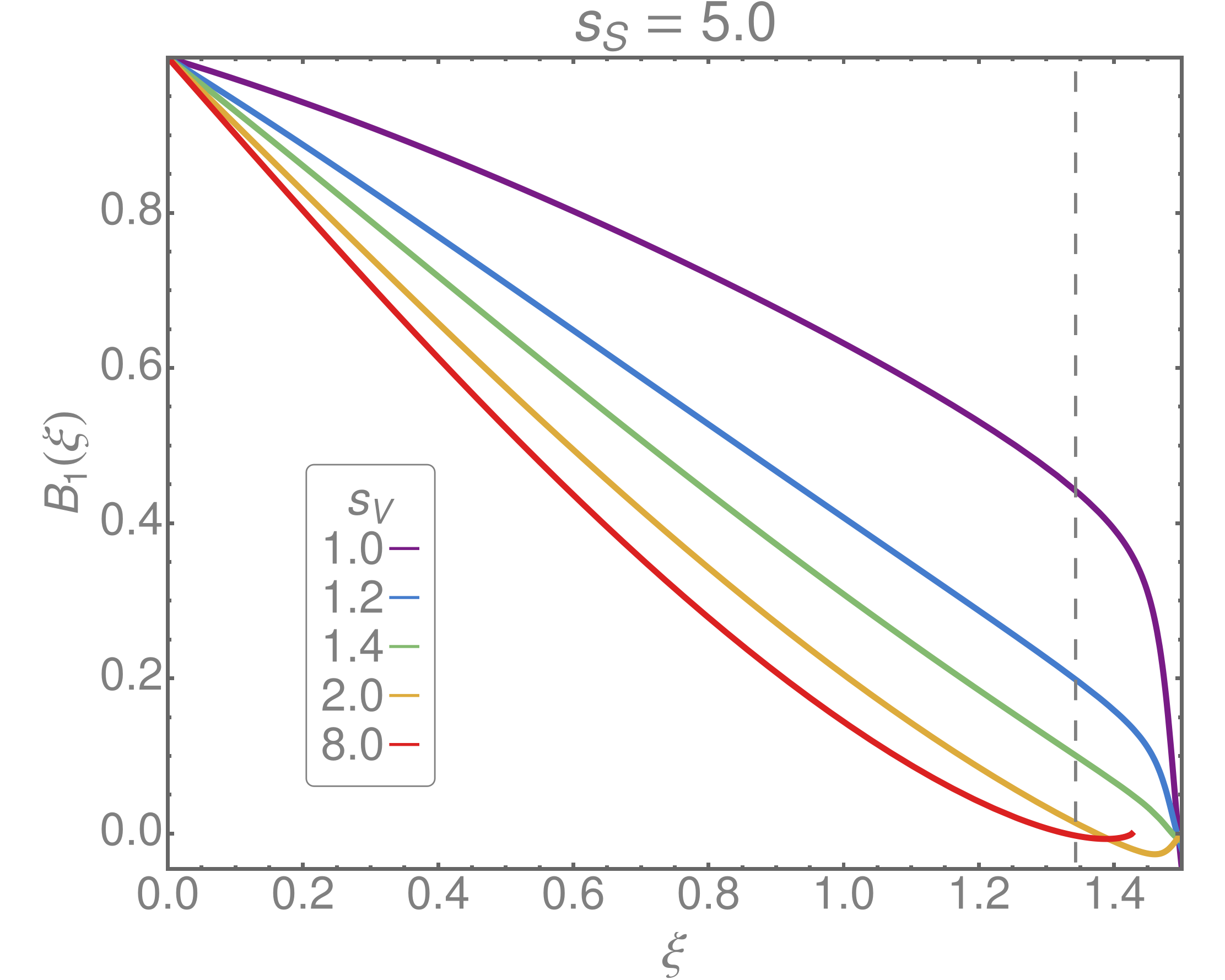}~~~
    \includegraphics[width=0.5\linewidth]{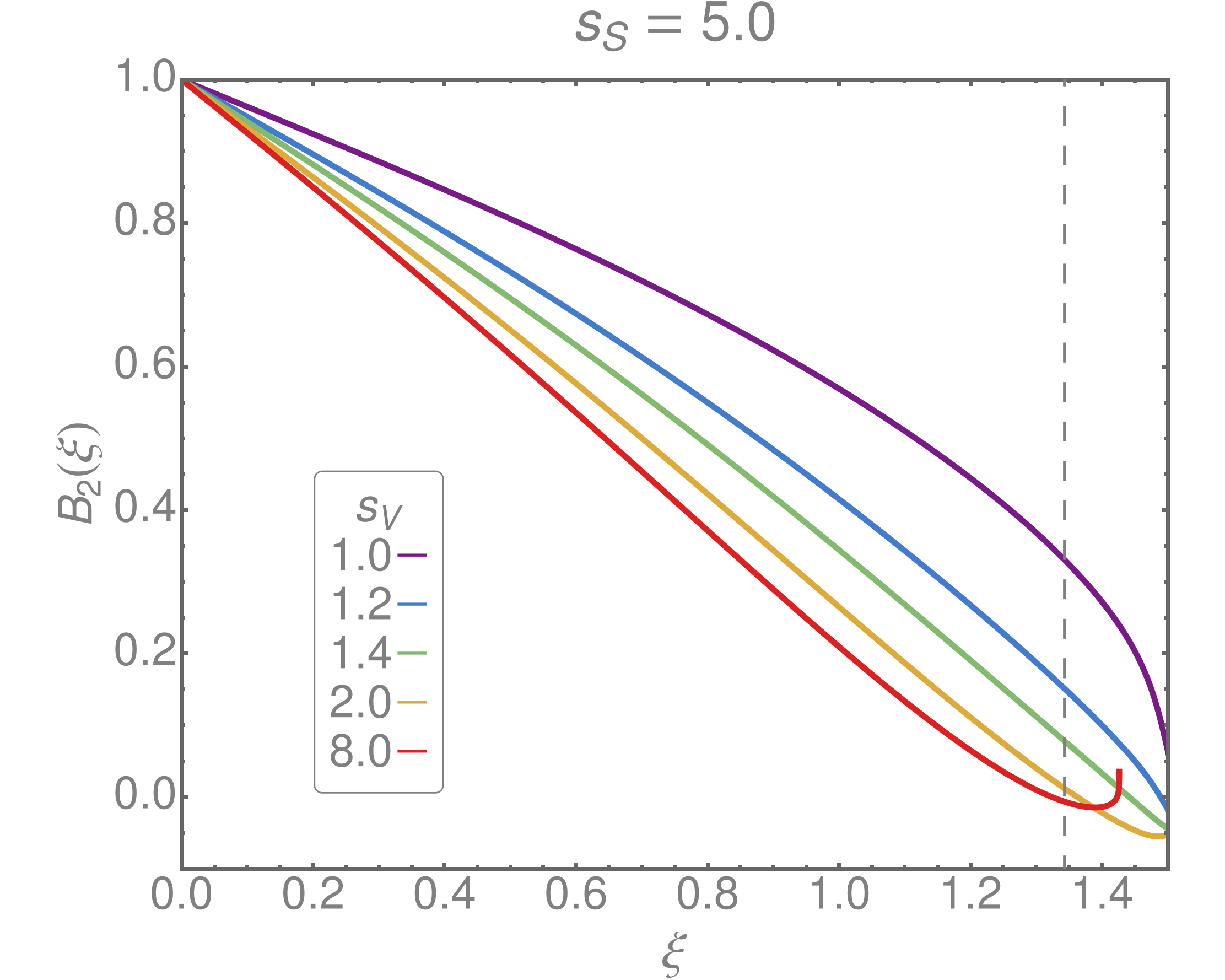}
  \caption{Functions $B_1(\xi)$ and $B_2(\xi)$ describing \ae ther perturbation (\ref{bdipole}) in the rest frame of a slowly moving Schwarzschild black hole for the scalar speed $s_S=5$ and several values of the vector speed~$s_V$.   
  When $s_V$ exceeds $2$, the perturbation quickly approaches the universal shape indistinguishable from the red curves. The dashed vertical line shows the location of the universal horizon of the unperturbed black hole $\xi_\star$ which is independent of $s_V$.}
  \label{fig:B1}
  \end{figure}
\begin{figure}[h!]
  \centering
  % row 1
  \hspace{-0.5cm}
    \includegraphics[width=0.5\linewidth]{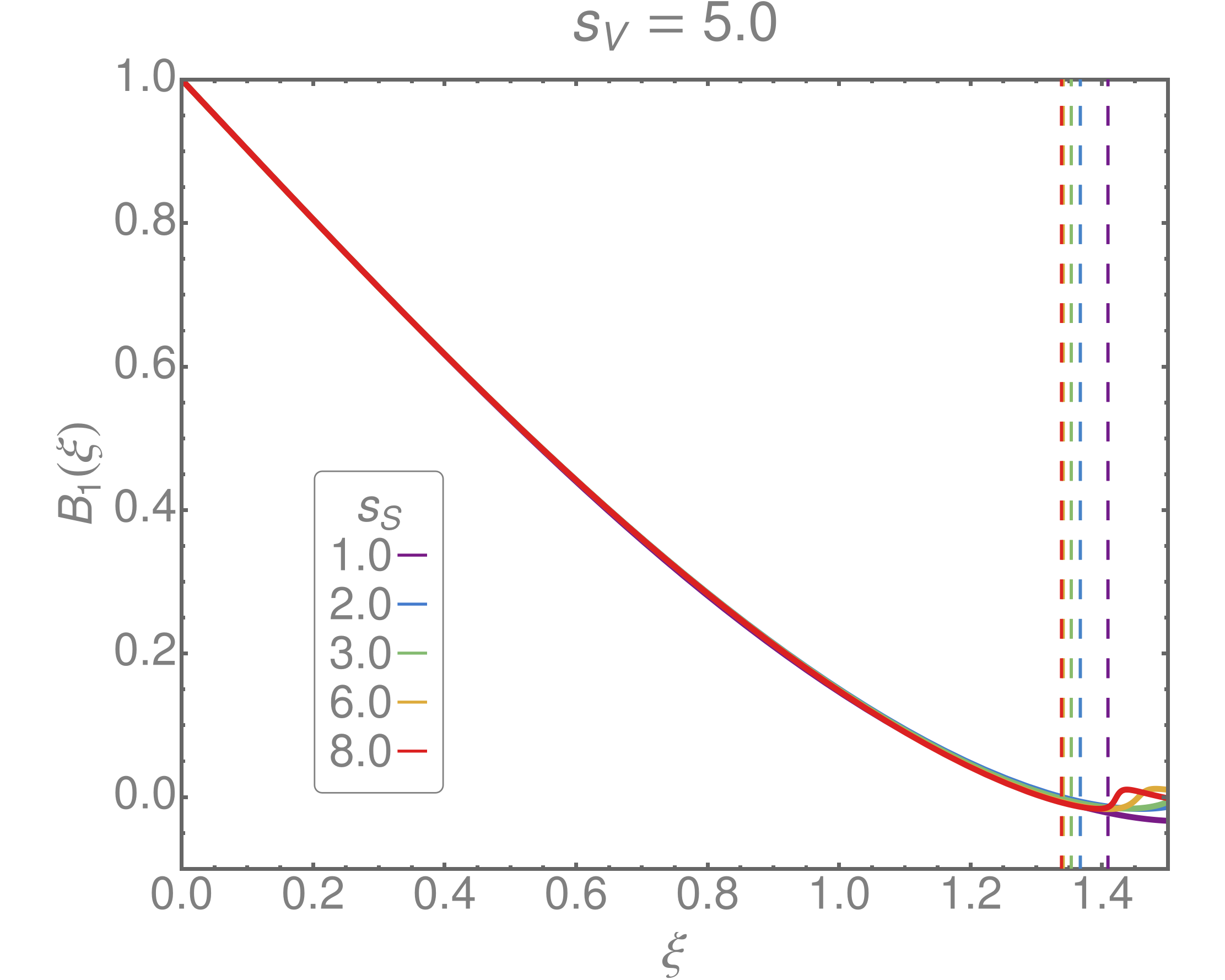}~~~
    \includegraphics[width=0.5\linewidth]{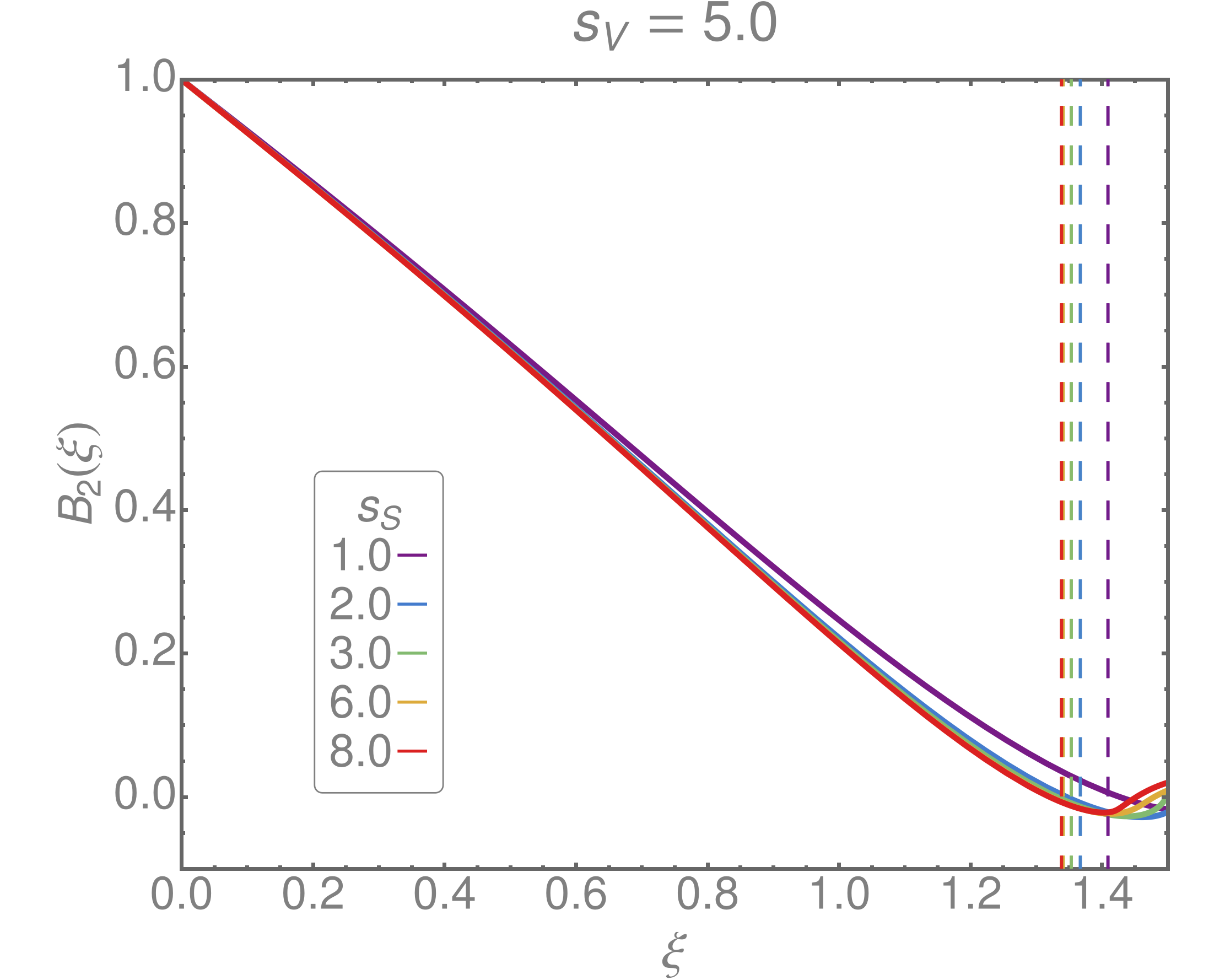}
  \caption{Same as Fig.~\ref{fig:B1}, but now for fixed $s_V=5$ and varying $s_S$. The dependence on $s_S$ is weak, so that the curves practically coincide. The universal horizon of the background solution $\xi_\star$ for each value of $s_S$ is shown with the same color as the corresponding perturbation.} \label{fig:B2}
  \end{figure}

Examples of numerical solutions are shown in Figs.~\ref{fig:B1} and \ref{fig:B2}. We see that the curves are regular and rather featureless outside of the universal horizon of the unperturbed solutions $\xi_\star$ shown by the dashed vertical lines. They approach universal limiting curves when either $s_V$ or $s_S$ increases. This approach happens faster for the variation of $s_S$, where the solutions are practically indistinguishable at all $s_S\geq 1$. We will study the limit of large excitation speeds analytically in Sec.~\ref{sec:largespeeds}.

The behavior at $\xi>\xi_\star$ is more varied and depends on the precise values of $s_S$ and $s_V$. In particular, for large $s_V$, a singularity appears at the inner helicity-1 horizon $\xi_{V,{\rm inner}}>\xi_\star$ which does not allow one to extend the solution past this value. This singular behavior is visible in the red curves in Fig.~\ref{fig:B1}. Its physical meaning is unclear. We stress, however, that it is causally disconnected from infinity by the outer horizons, and does not affect calculation of the black hole sensitivities.

\begin{figure}[h!]
  \centering
  % row 1
  \hspace{-0.5cm}
    \includegraphics[width=0.5\linewidth]{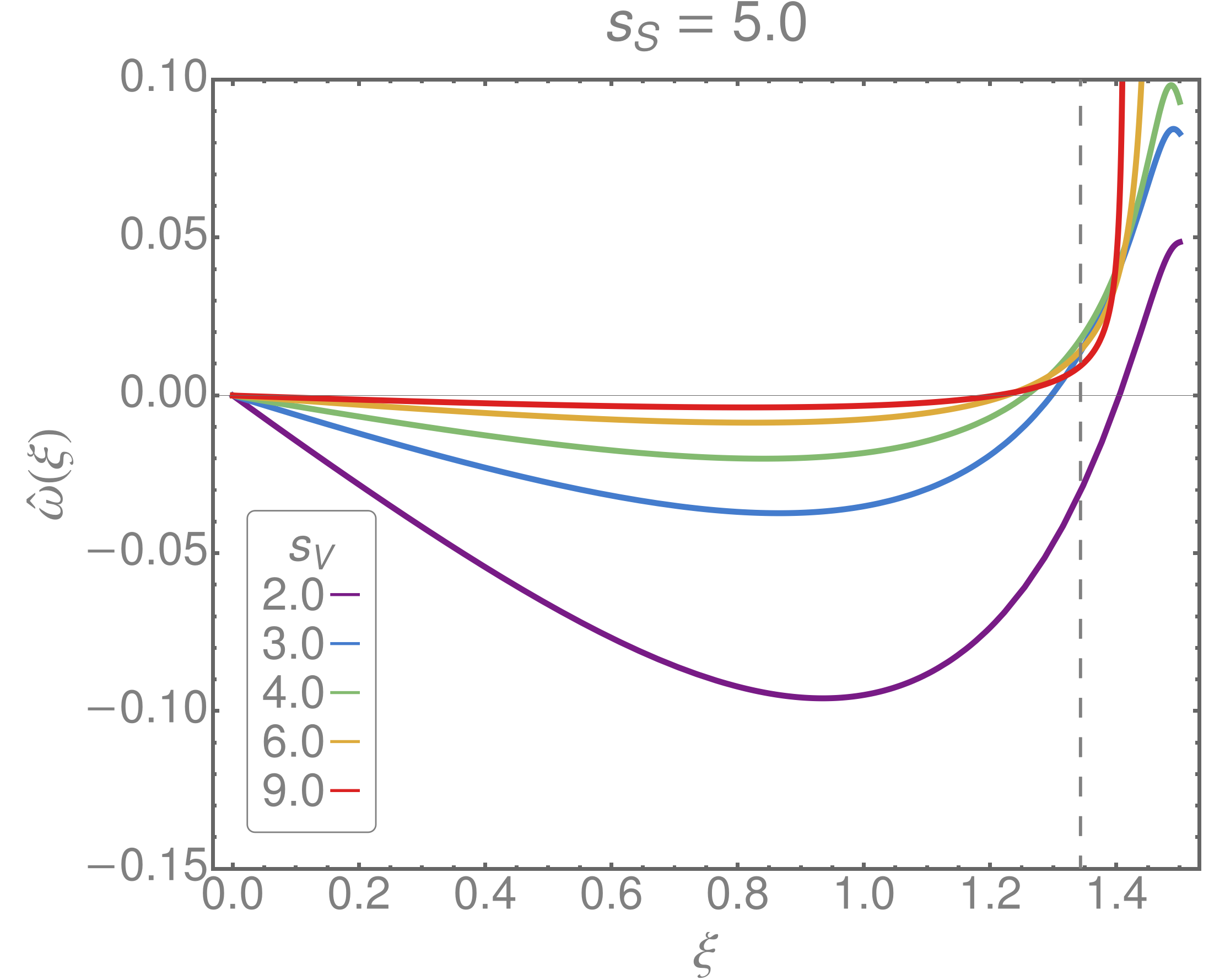}~~~
    \includegraphics[width=0.5\linewidth]{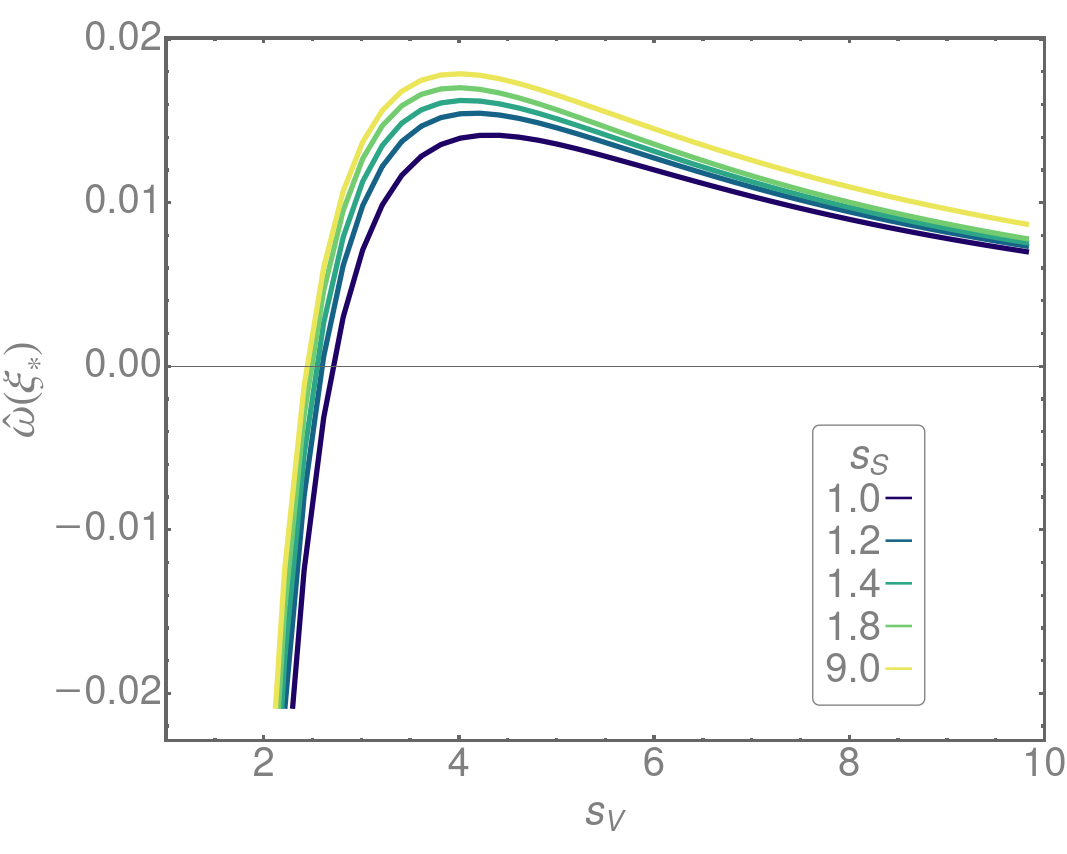}
  \caption{\textit{Left:} The function $\hat\omega(\xi)$ describing the twist tensor of the perturbed \ae ther, defined in Eq.~(\ref{omegalin}). Vertical dashed line shows the location of the universal horizon for the background solution. \textit{Right:} The value of $\hat{\omega}$ evaluated at the universal horizon of the background solution~$\xi_\star$.} \label{fig:omega}
  \end{figure}

It is instructive to study if the moving BH possesses a universal horizon, like the static one, or not. If the universal horizon persists, it represents a surface orthogonal to the perturbed \ae ther field, implying that on the universal horizon the \ae ther twist necessarily vanishes (see the discussion in \cite{Adam:2021vsk,Barausse:2015frm}). From Eq.~(\ref{omegalin}) we see that the non-trivial components of the twist are proportional to a single function $\hat\omega(\xi)$. In the left panel of Fig.~\ref{fig:omega} we plot this function for several solutions differing by the value of $s_V$. We see that the twist exhibits non-monotonic behavior: it decreases from $\hat\omega=0$ at spatial infinity ($\xi=0$) to negative values, but then turns around, crosses zero and becomes positive. Depending on $s_V$, the zero crossing happens at $\xi$ bigger or smaller than~$\xi_\star$. 

The right panel of Fig.~\ref{fig:omega} shows the twist function $\hat\omega$ evaluated at $\xi_\star$. In general, $\hat\omega(\xi_\star)$ is non-vanishing, implying that it is impossible to choose a surface in the vicinity of $\xi=\xi_\star$ that would be orthogonal to the perturbed \ae ther. Thus, the motion of the BH generically destroys the universal horizon. 
We note, however, that for any value of the scalar speed $s_S$, there is a value of $s_V$, such that $\hat\omega(\xi_\star)$ does vanish. For this particular fine-tuned choice of $s_S$, $s_V$ the universal horizon appears to persist even for moving BHs.

The universal horizon also persists in the khrono-metric limit $s_V\to\infty$. Indeed, we observe in Fig.~\ref{fig:omega} that the overall magnitude of the twist decreases with growing $s_V$ everywhere at $\xi\leq\xi_\star$. The limit at $\xi_\star$, however, is subtle. For large $s_V$, the twist diverges inside $\xi_\star$ at the inner vector horizon $\xi_{V,{\rm inner}}$. The latter approaches $\xi_\star$ from above at $s_V\to\infty$, making the limit of the solution at $\xi_\star$ non-analytic. We will further discuss this point in Sec.~\ref{sec:largespeeds}.

\section{Sensitivity} \label{sec:sens}
\subsection{Point particle effective theory}
\label{ssec:EFT}

In realistic situations, BH does not exist in isolation but interacts with other bodies and fields. When the distance between the bodies or the wavelength of the external field is large compared to the BH size, many aspects of the interaction can be captured by the effective point particle description. In this framework, the BH is described by an effective worldline action carrying operators coupling the worldline degrees of freedom to other fields. The couplings obey the symmetries of the microscopic Lagrangian and are organized using the derivative expansion \cite{Goldberger:2004jt,Porto:2016pyg, Goldberger:2022ebt}.

We are interested in the coupling of non-rotating BH to \ae ther. The corresponding point particle action must be generally covariant, as the consequence of the general covariance of the underlying \AE-theory. It is thus constructed using Lorentz-invariant contractions of the \ae ther vector $u^\mu$, particle four-velocity $v^\mu$, and their covariant derivatives. At the lowest order in derivatives, the effective action has the form \cite{Foster:2007gr}, 
\begin{equation}
\label{ppaction}
S_{pp} = -m \int \text{d}\tau\, F(\gamma)\;,
\end{equation}
where $\tau$ is the particle's proper time and 
$\gamma \equiv u_\mu v^\mu$ describes the relative boost between the particle and \ae ther. The function $F(\gamma)$ is normalized by the condition $F(1)=1$, so that $m$ coincides with the mass of the particle at rest with respect to the \ae ther, but is otherwise arbitrary. For slow relative motion, it can be expanded around $\gamma=1$ and (\ref{ppaction}) becomes
\begin{equation}
\label{ppaction1}
S_{pp} = - m \int \text{d}\tau \left(1+\sigma (1-\gamma) + \frac{\sigma'}{2}(1-\gamma)^2 + \ldots\right)\;.
\end{equation}
The first term gives the usual relativistic point particle action. In the rest-frame of the \ae ther $1-\gamma=-v^2/2+{\cal O}(v^4)$, so the subsequent terms are ordered in powers of the relative particle-\ae ther velocity. The coefficients $\sigma$, $\sigma'$, etc. are called `sensitivities' \cite{Foster:2007gr}. In what follows, we focus on the first sensitivity $\sigma$ which controls the dominant interaction at $v\ll 1$. 

Our goal is to compute $\sigma$ from the slowly moving BH solutions found in the previous section. To this aim, we will find the static perturbation of the \ae ther produced by the point particle with the action (\ref{ppaction1}) in the particle rest frame. This \ae ther perturbation will depend on $\sigma$ as a free parameter. We will then identify this configuration with the asymptotics of the full BH solution which we have shown in Sec.~\ref{sec:movingbh} to be unique. This will fix $\sigma$ in terms of the \ae ther couplings $c_\vartheta$, $c_\omega$, $c_a$, $c_\sigma$.  

In the absence of particle, the metric and \ae ther in the reference frame moving with velocity ${\bf v}$ relative to the \ae ther are,
\be
\label{ppbckg}
\tilde g_{\mu\nu}=\eta_{\mu\nu}~,~~~~ \tilde u^\mu\d_\mu=\frac{1}{\sqrt{1-v^2}}(\d_t-v^i\d_i)\;,
\ee
where $\eta_{\mu\nu}$ is the Minkowski metric. Latin indices from the middle of the alphabet will stand for spatial directions and we will be summing over them with the trivial metric $\delta_{ij}$, irrespective of their positions. We will nevertheless distinguish upper and lower indices, in that the corresponding quantities differ by sign, e.g. $\tilde u_i=-\tilde u^i=v^i/\sqrt{1-v^2}$. 

Adding a particle perturbs the metric and \ae ther, so we write
\begin{equation}
\label{pppert}
g_{\mu\nu}=\eta_{\mu\nu} +h_{\mu\nu}\;,~~~~u^\mu=\tilde u^\mu+d^\mu\;.
\end{equation}
Since we still assume the decoupling limit, the metric perturbation is known and is written in terms of the Newtonian potential $\Phi({\bf x})$,
\begin{equation}
\label{hpert}
h_{\mu\nu} \text{d}x^\mu \text{d}x^\nu = 2 \Phi({\bf x}) 
\big(\text{d}t^2 + \text{d}{\bf x}^2\big)\;, ~~~~ \Phi({\bf x}) =  - \frac{Gm}{|x|}\;.
\end{equation}
Our task is to find the \ae ther perturbation $d^\mu$. It is important to note that the perturbative expansion we are using here differs from the expansion in Sec.~\ref{sec:movingbh}.\footnote{To emphasize this fact, we use different notations from Sec.~\ref{sec:movingbh} for the background and perturbation of the \ae ther vector.} 
There, we treated the problem to all orders in the gravitational coupling $G$ and expanded in small velocity $v^i$. In the point particle case, we expand in $G$ (more precisely, in $Gm$) and, in principle, can work to all orders in $v^i$. This is why the asymptotic tilt of the \ae ther has been included in the leading-order expression (\ref{ppbckg}). In practice, we will also use linear approximation in $v^i$, since this greatly simplifies the formulas and is sufficient to extract the sensitivity $\sigma$.  

The most economic way to derive the equations for $d^\mu$ is again to use the action principle. The point particle action, which provides the source for $d^\mu$, needs to be expanded only to linear order. Using the unit-norm constraint (\ref{unitnorm}), we find 
\be
\label{ppdt}
d_t=\Phi-v^i d^i+{\cal O}(v^2)\;.
\ee
Inserting this into (\ref{ppaction1}) gives the source term,
\begin{equation}
\label{ppsource}
S_{pp}^{(1)} =  -m \sigma \int \dd t\, v^i d^i\;,
\end{equation}
where the field $d^i$ is evaluated at ${\bf x}=0$.
From the \ae ther action, we need quadratic and linear in $d^i$ terms. Since all the contributions in (\ref{e:AEAction2}) vanish in the background (\ref{ppbckg}), it is sufficient to compute the expansion, acceleration and twist up to terms linear in $d^i$ and $\Phi$. Discarding also higher than linear dependence on $v^i$, we obtain
\bseq
\label{ppperts}
\begin{gather}
\vartheta^{(1)}=\d_i d^i+2 v^i\d_i\Phi\;, \\
a_t^{(1)}=-v^i\d_i\Phi\;,~~~~a_i^{(1)}=-\d_i\Phi+v^j\d_j d^i\;,\\
\omega_{ti}^{(1)}=\frac{1}{2}v^j\d_i d^j-\frac{1}{2}v^j\d_j d^i\;,~~~~
\omega_{ij}^{(1)}=-\frac{1}{2}\d_i d^j+\frac{1}{2}\d_j d^i
+\frac{3}{2} v^i\d_j\Phi-\frac{3}{2}v^j\d_i\Phi\;.
\end{gather}
\eseq
Squaring these expressions and adding perturbation of the last term in (\ref{e:AEAction2}),
we arrive at the quadratic action,
\begin{align}
S_{AE}^{(2)} =
\frac{-1}{16\pi G} \int \dd[4]{x} \bigg[&
-\frac{c_\omega}{2} d^i\Delta d^i
+\bigg(\frac{c_\omega}{2}-\frac{c_\vartheta}{3}\bigg)
d^i \partial_i\partial_j d^j+(4c_\sigma-3c_\omega)d^i v^i \,\Delta \Phi
\notag\\
&+\bigg(3c_\omega-\frac{4c_\vartheta}{3}-2c_a\bigg) d^i v^j \,\partial_i \partial_j \Phi
\bigg]\;.
\label{pppertact}
\end{align}
Here $\Delta\equiv \d_i\d_i$ denotes the Laplacian, and we have used 
the linear perturbation of the Ricci tensor, $R_{tt}^{(1)}=\Delta \Phi$, $R_{ij}^{(1)}=\Delta \Phi\, \delta_{ij}$.

Equation for $d^i$ is obtained by varying the sum of (\ref{pppertact}), (\ref{ppsource}) with respect to $d^i$,
\be
\label{prefourierEOM}
\begin{split}
-c_\omega \Delta d^i+\left(c_\omega\!-\!\frac{2}{3}c_\vartheta\right)
\partial_i\partial_j d^j
+(4c_\sigma\!-\!3c_\omega)\, v^i \Delta \Phi
+\left(3c_\omega\!-\!\frac43 c_\vartheta\!-\!2c_a\right) v^j\partial_i \partial_j \Phi&\\
+16\pi G m \sigma v^i \delta({\bf x}) &= 0\;.
\end{split}
\ee
To solve it, we switch to the Fourier space and split $d^i({\bf k})$ into the longitudinal and transverse parts,\footnote{With some abuse of notations, we use the same letter $d^i$ for the \ae ther perturbation and its Fourier transform.}
\be
\label{ddecomp}
d^i({\bf k})=d_\perp^i({\bf k})+k^id_\parallel ({\bf k})\;,~~~~k^id_\perp^i({\bf k})=0\;.
\ee
The equation then takes the form,
\begin{align}
\label{equationtoinverse}
-c_\omega k^2 d_\perp^i-\frac{2}{3}c_\vartheta k^ik^2d_\parallel=
4\pi G m \bigg[(4c_\sigma-3c_\omega+4\sigma)v^i
+\bigg(3c_\omega-\frac{4}{3}c_\vartheta-2c_a\bigg)\frac{k^ik^j}{k^2}v^j\bigg]\;.
\end{align}
Contracting it with $k^i$, we find the longitudinal component $d_\parallel$ which, substituted back in (\ref{equationtoinverse}), determines the transverse part $d_\perp^i$. Combining them together, we obtain the solution,
\be
\label{dksolut}
d^i({\bf k})=-\frac{4\pi Gm}{k^2}\bigg[
\frac{(6c_\sigma-2c_\vartheta-3c_a+6\sigma)}{c_\vartheta}\,
\frac{k^ik^j}{k^2}
+\frac{(4c_\sigma-3c_\omega+4\sigma)}{c_\omega}
\bigg(\delta^{ij}-\frac{k^ik^j}{k^2}\bigg)
\bigg] v^j\;.
\ee
To transform it back to coordinate space, we use the 
inverse Fourier maps:
\be
\label{Fouriermaps}
-\frac{4\pi k^ik^j}{k^4}\mapsto 
-\frac{1}{2|x|}\bigg(\delta^{ij}-\frac{x^ix^j}{|x|^2}\bigg)\;,~~~~
-\frac{4\pi}{k^2} \bigg(\delta^{ij}-\frac{k^ik^j}{k^2}\bigg)
\mapsto 
-\frac{1}{2|x|}\bigg(\delta^{ij}+\frac{x^ix^j}{|x|^2}\bigg)\;.
\ee
Applying them to (\ref{dksolut}), we arrive at
\begin{equation}
\label{fullUwithPert}
d^i({\bf x}) = \frac{Gm\,v^i}{|x|}\bigg[\frac{5}{2}-\frac{2\hat\sigma}{c_\omega}+\frac{3c_a}{2c_\vartheta}-\frac{3\hat \sigma}{c_\vartheta}\bigg]
+\frac{Gm\, x^i({\bf x}\!\cdot\!{\bf v})}{|x|^3}
\bigg[\frac{1}{2}-\frac{2\hat\sigma}{c_\omega}-\frac{3c_a}{2c_\vartheta}+\frac{3\hat\sigma}{c_\vartheta}\bigg],
\end{equation}
where we have introduced the notation
\be
\label{newsigma}
\hat \sigma\equiv\sigma+c_\sigma\;.
\ee
Note that since $c_\sigma$ is observationally constrained to be very small (see Sec.~\ref{ssec:Lagr}), the shifted sensitivity $\hat\sigma$ is essentially equal to $\sigma$.
The expression (\ref{fullUwithPert}) is the final result of this subsection. We now discuss its matching to the full BH solution.

\subsection{Matching effective theory to black hole}

The formula (\ref{fullUwithPert}) is expected to describe the behavior of the \ae ther far from the BH. Thus it must agree with the large-distance asymptotics of the solution found in Sec.~\ref{sec:movingbh}. In more detail, comparing the forms (\ref{aetherpert}) and (\ref{pppert}), we require
\be
\label{matching}
b^i\simeq \tilde u^i+d^i\qquad \text{at}~~r\to\infty\;.
\ee
The matching is somewhat complicated by the fact that the solutions of Secs.~\ref{sec:movingbh} and \ref{ssec:EFT} are found in different spatial coordinate frames: Sec.~\ref{sec:movingbh} uses the Schwarzschild coordinates, whereas Sec.~\ref{ssec:EFT} employs the Cartesian frame. Of course, it is straightforward to cast the vector (\ref{pppert}) into polar coordinates, in which the metric reads 
\be
\label{polarmetr}
\dd{\bf s}^2=(1+2\Phi)\dd t^2-(1-2\Phi)\big(\dd\tilde r^2+\tilde r^2(\dd\theta^2+\sin^2 \theta\, \dd\phi^2)\big)\;.
\ee
The angular variables of this and Schwarzschild frames coincide. However, the radial variables differ. This can be seen by comparing the angular parts of the two metrics:
\be
\label{angmetr}
r^2\dd\Omega^2=(1-2\Phi)\,\tilde r^2\dd\Omega^2\;,
\ee
which implies, to first order in $\frac{Gm}{\tilde r}$, the relation
\be
\label{rrho}
r=\tilde r+Gm\;.
\ee
With this at hand, we can equate
\be
\label{matching1}
b^r\simeq \tilde u^{\tilde r}+d^{\tilde r}~,~~~~b^\theta\simeq \tilde u^\theta+d^\theta\;.
\ee

The l.h.s. of the first relation is given by (\ref{bhatB}) and at large $r$ takes the form,
\be
\label{brasymp}
b^r=-v\cos\theta+\big(1-2B'_1(0)\big)\frac{Gm v}{r}\cos\theta+{\cal O}(1/r^2)\;,
\ee
where $B'_1(0)$ is the derivative of the function $B_1(\xi)$ at $\xi=0$. In deriving this expression we have used the identity (\ref{Uder0}) and restored the dimensionful units by substituting $\xi=2Gm/r$. On the other hand, the r.h.s. of the first equation in (\ref{matching1}) reads,
\be
\label{ppurho}
\tilde u^{\tilde r}+d^{\tilde r}=\frac{x^i}{\tilde r}(\tilde u^i+d^i)
=-v\cos\theta+\bigg(3-\frac{4\hat\sigma}{c_\omega}\bigg) \frac{Gmv}{\tilde r}\cos\theta\;.
\ee
Comparing this to Eq.~(\ref{brasymp}), we read off the sensitivity,
\begin{equation}
\label{sensitivityEqn}
\hat\sigma = c_a s_V^2 \big(1+B_1'(0)\big)\;,
\end{equation}
where we have used the expression (\ref{speeddefns}) for $s_V^2$.

It may appear from Eq.~(\ref{sensitivityEqn}) that the sensitivity diverges in the khrono-metric limit $s_V\to\infty$, at $c_a$, $s_S$ fixed. This, however, cannot happen because the sensitivity for the khrono-metric theory has been calculated and is finite \cite{Kovachik:2023fos}. To see that the sensitivity remains finite at $s_V\to\infty$, 
we can use the second matching condition (\ref{matching1}). From Eq.~(\ref{bdipole}), its l.h.s. is 
\be
\label{bthetasymp}
b^\theta=\frac{vB_2}{r}\sin\theta=\frac{v}{r}\sin\theta+2B_2'(0) \frac{Gmv}{r^2}\sin\theta+{\cal O}(1/r^3)\;.
\ee
Whereas the r.h.s. has the form,
\be
\label{dthetmatch}
\tilde u^\theta+d^\theta=\frac{v}{\tilde r}\sin\theta
-\bigg(\frac{5}{2}-\frac{2\hat\sigma}{c_\omega}+\frac{3c_a}{2c_\vartheta}-\frac{3\hat\sigma}{c_\vartheta}\bigg) \frac{Gmv}{\tilde r^2}\sin\theta\;.
\ee
In comparing these two expressions, we must take into account the relation between the radial coordinates (\ref{rrho}). In this way we obtain,
\be
\hat\sigma =c_a \,\frac{s_V^2}{2(s_S^2+s_V^2)}\big(1+3s_S^2 + 4s_S^2 B_2'(0)\big).
\label{sigmaInB2}
\ee
This expression is more cumbersome than Eq.~(\ref{sensitivityEqn}) but it has the advantage of being manifestly regular in the khrono-metric limit (recall that $B_2'(0)$ remains finite at $s_V\to\infty$, as discussed in Sec.~\ref{sec:singular points}). 

Equations (\ref{sensitivityEqn}), (\ref{sigmaInB2}) provide different expressions for the same quantity. Are they compatible with each other? The answer is yes. The analysis of Appendix~\ref{sec:logtermneed} reveals that for any two functions $B_1(\xi)$, $B_2(\xi)$ solving the aether equations (\ref{perteqs}) and regular at the origin, their derivatives are related by the condition
\be
\label{Bpr2Bpr1}
B_2'(0) = -\frac{1}{4}-\frac{1}{4s_S^2}+\frac{s_V^2}{2s_S^2}+\frac{s_S^2+s_V^2}{2s_S^2} B_1'(0)\;,
\ee
which ensures that Eqs.~(\ref{sensitivityEqn}) and (\ref{sigmaInB2}) coincide. 
This agreement provides a non-trivial check of our calculations. 
Of course, to determine either $B_1'(0)$ or $B_2'(0)$ and hence the sensitivity, one needs to solve Eqs.~(\ref{perteqs}) globally, imposing regularity not only at $\xi=0$, but also at the horizons $\xi_S$ and $\xi_V$. 

\begin{figure}[t]
    \centering
    \hspace{-0.7cm}
   \begin{subfigure}[t]{0.51\textwidth}
    \centering
    \includegraphics[width=\linewidth]{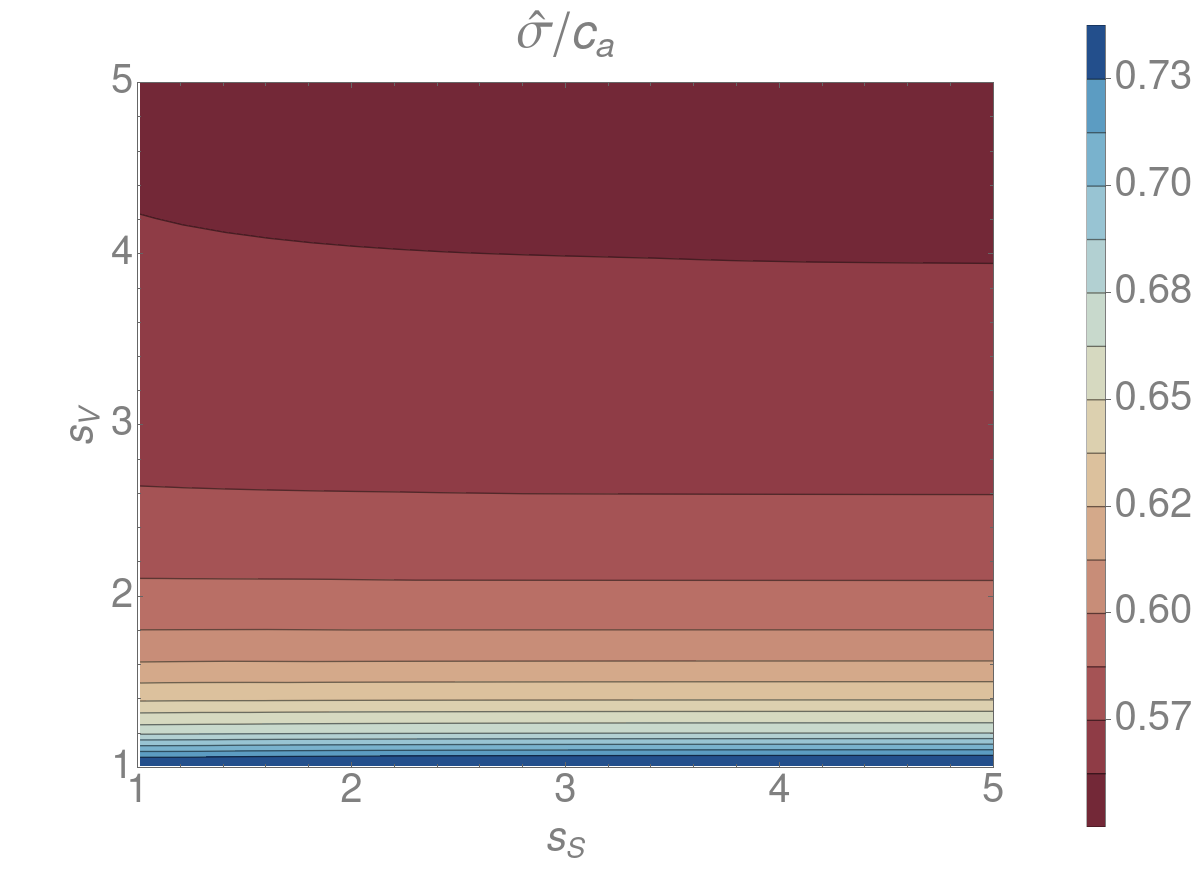}
    \end{subfigure}~~~~~~~
   \begin{subfigure}[t]{0.5\textwidth}
   \includegraphics[width=.85\linewidth]{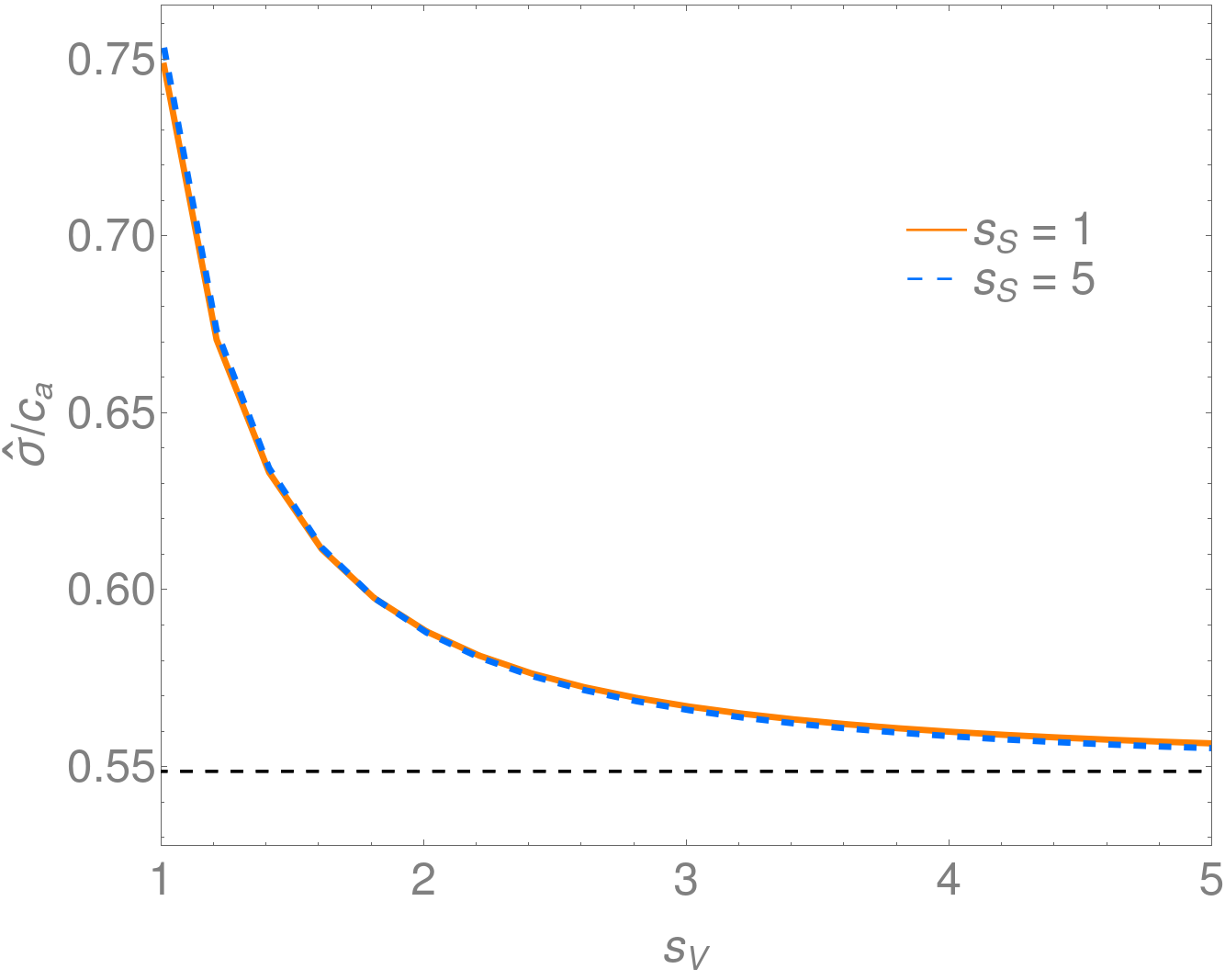}
%    \label{fig:sensitivity values slice}
\end{subfigure}
\caption{Shifted sensitivity $\hat \sigma$, Eq.~(\ref{newsigma}), normalized to the \ae ther coupling $c_a$. {\it Left:} Contour plot in the $(s_S,s_V)$ plane.
{\it Right:}  $s_V$-dependence at two fixed values of $s_S$. Note that the curves almost coincide. Black dashed line shows the asymptotic value (\ref{khronosens}). 
}
  \label{fig:sensitivity values}
\end{figure}

In Fig.~\ref{fig:sensitivity values} we show the ratio $\hat\sigma/c_a$ extracted from our numerical solutions. 
This ratio is a function of the scalar and vector mode speeds $s_S$ and $s_V$
which are sampled from a uniform grid with step size $0.25$ 
in the range of $1\le s_S, s_V \le 5$.
We observe that $\hat\sigma/c_a$ depends very weakly on $s_S$. Similar weak dependence of the sensitivity on the scalar speed was observed in the khrono-metric case \cite{Kovachik:2023fos}.
The dependence on $s_V$ is more pronounced: the ratio monotonically decreases from $\hat\sigma/c_a\approx 0.75$ at $s_V=1$ to $\hat\sigma/c_a\approx 0.55$ at $s_V\to \infty.$

\begin{figure}
    \centering
    \hspace{-0.7cm}
    \begin{subfigure}[t]{0.5\textwidth}
    \centering
    \vspace{-6.1cm}
    \includegraphics[width=\linewidth]{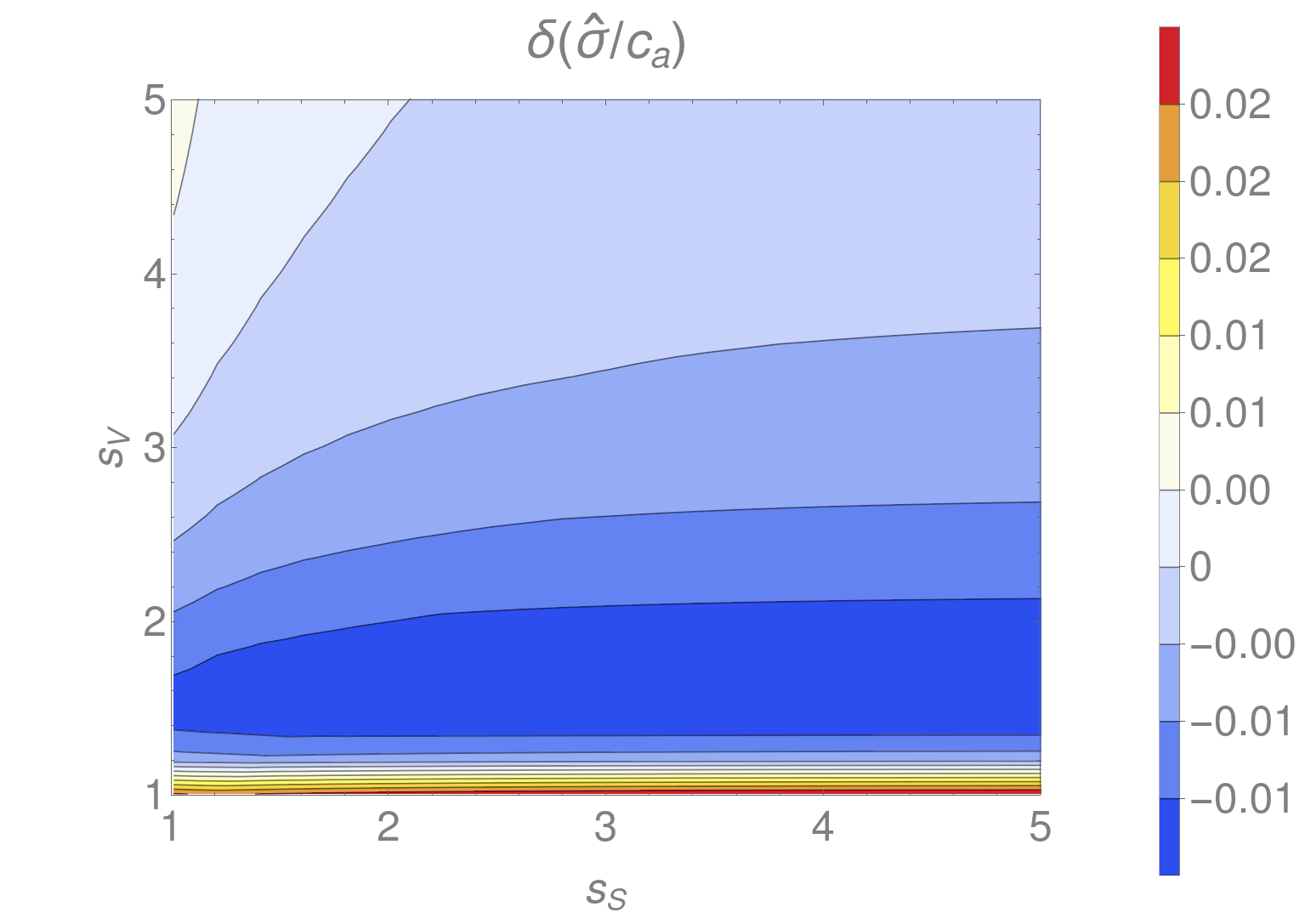}
    \end{subfigure}~~~~~~~
    \begin{subfigure}[t]{0.5\textwidth}
    \includegraphics[width=.85\linewidth]{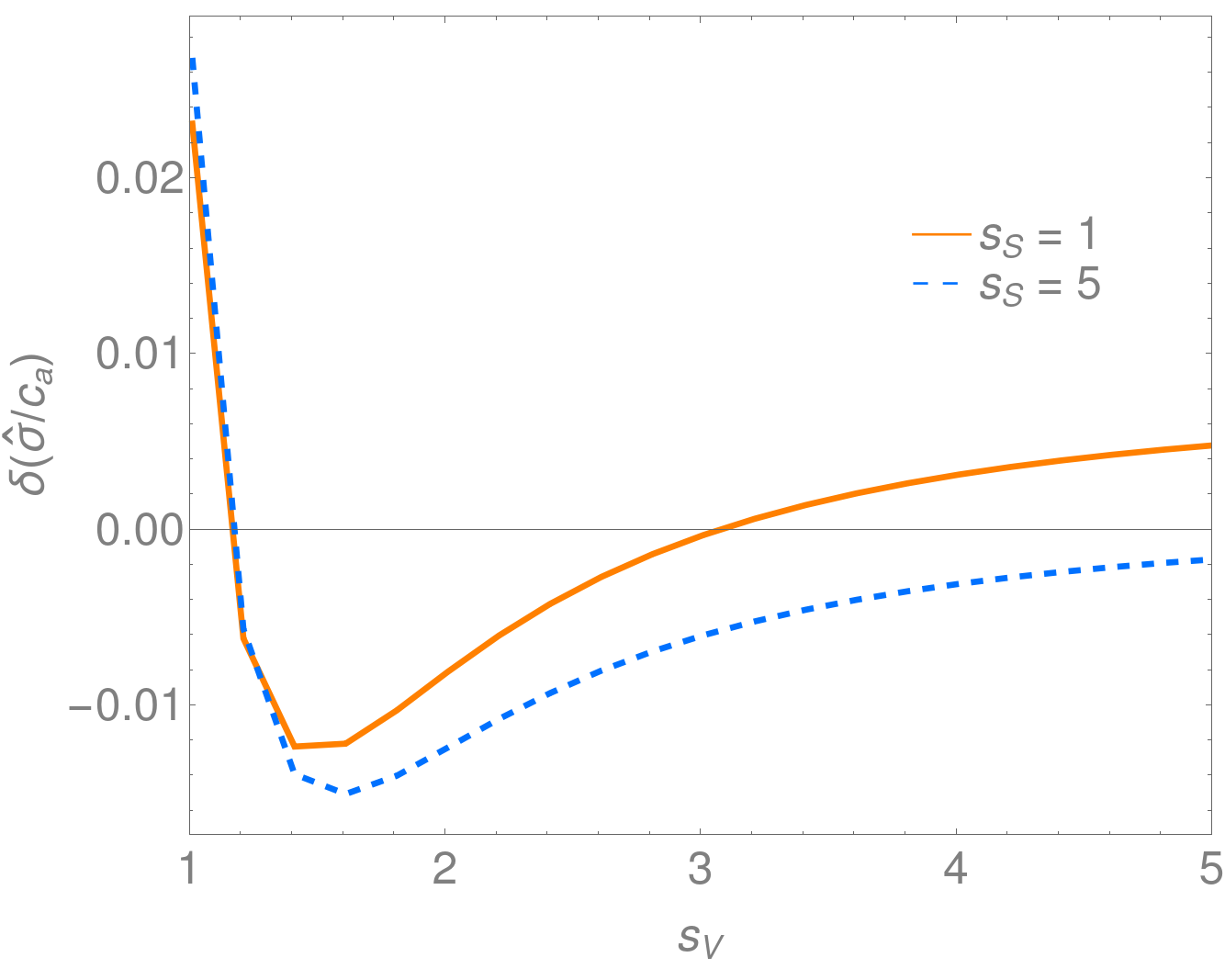}
    \end{subfigure}
     \caption{Relative difference (\ref{residualdef}) between the numerical results for $\hat\sigma/c_a$ and the analytic fit (\ref{sigmaSeriesDef}). {\it Left:} Contour plot in the $(s_S,s_V)$ plane.
{\it Right:} $s_V$-dependence at two fixed values of $s_S$. 
}
    \label{fig:sensitivity difference values}
\end{figure}

In the next section, we show that in the limit $s_S, s_V\to\infty$ the ratio acquires the same value as in the khrono-metric theory and can be computed rather precisely,
\be
\label{khronosens}
\left(\frac{\hat\sigma}{c_a}\right)_\infty=0.5488\;.
\ee
Corrections to this value are expected to scale in powers of $s_S^{-2}$ and $s_V^{-2}$. 
Numerically, we find that the data in the range $1\leq s_S,s_V\leq 10$ are well fitted by the expression
\be
\label{sigmaSeriesDef}
\frac{\hat\sigma}{c_a} \simeq \left(\frac{\hat\sigma}{c_a}\right)_\infty+ \frac{A_1}{s_S^2} + \frac{A_2}{s_V^2}\;,~~~~~~~~ A_1 = -0.002, \ \ A_2 = 0.188\;.
\ee
The relative difference
between the exact numerical results and the fit (\ref{sigmaSeriesDef}),
\be
\label{residualdef}
\delta (\hat\sigma/c_a)\equiv \frac{\hat\sigma/c_a-(\hat\sigma/c_a)_\text{fit}}{\hat\sigma/c_a}\;,
\ee
is shown in Fig.~\ref{fig:sensitivity difference values}. We see that it does not exceed a few per cent level.

Note that the sensitivity (\ref{sensitivityEqn}), (\ref{sigmaInB2}) is independent of the BH mass. This could be expected, since the BH mass (or rather its Schwarzschild radius) is the only dimensionful parameter in the problem and thus cannot affect a dimensionless quantity, such as the sensitivity. Note also that the sensitivity is proportional, with order-one coefficient, to the coupling $c_a$, which is constrained to be small, yet not as small as $c_\sigma$; see Sec.~\ref{ssec:Lagr}. Investigation of the phenomenological impact of such sensitivity beyond the scope of this paper. Here we only note that it affects the BH metric at order ${\cal O}(c_a)$ through modification of the PPN parameters $\alpha_1^{\rm PPN}$, $\alpha_2^{\rm PPN}$ \cite{Yagi:2013ava}. In the limit $c_\sigma\ll c_a\ll 1$, the expressions for these parameters including the sensitivity contribution are, 
\begin{align}
\label{PPNstrong}
\tilde \alpha_1^\text{PPN} = \alpha_1^\text{PPN} + 8\sigma\;,\qquad~~~
\tilde \alpha_2^\text{PPN} = \alpha_2^\text{PPN} + \left(4+ \frac{\alpha_1^\text{PPN} - 2 \alpha_2^\text{PPN}}{c_a} \right)\sigma\;,
\end{align}
where $\alpha_1^{\rm PPN}$, $\alpha_2^{\rm PPN}$ are the weak-field values 
(\ref{alpha12Def}).
Using the approximate formula (\ref{sigmaInB2}) we obtain,
\bseq
\label{PPNstrongfit}
\begin{align}
\tilde\alpha_1^\text{PPN} &\simeq c_a\left(0.39 - \frac{0.02}{s_S^2} + \frac{1.50}{s_V^2}\right)\;,\\
\tilde\alpha_2^\text{PPN} &\simeq c_a \left(1-\frac{1}{s_S^2}\right)
\left(0.049-\frac{0.002}{s_S^2} + \frac{0.188}{s_V^2}\right)\;.
\end{align}
\eseq
We observe that the proportionality coefficient between $\tilde\alpha_1^\text{PPN}$ and $c_a$ is order-one, whereas for $\tilde\alpha_2^\text{PPN}$ it is an order of magnitude smaller. The dependence on $s_V$ is quite significant, with the parameters at $s_V=1$ being a factor of five larger than at $s_V\to\infty$. 
Curiously, the parameters (\ref{PPNstrongfit}) have opposite sign, compared to their weak-field counterparts (\ref{alpha12Def}).
Note also that the second parameter $\tilde \alpha_2^\text{PPN}$ vanishes for $s_S=1$, similar to the weak-field value.  

\section{Khrono-metric limit}\label{sec:largespeeds}

To gain an analytic insight into the dependence of the sensitivity on the parameters $s_S$ and $s_V$, we study the slowly moving BH solution in the limit when these parameters are large.
We consider two ways of taking the limit. First, we take $s_V\to\infty$ with $s_S$ kept fixed. In this limit, the \ae ther must become hypersurface-orthogonal and we expect to recover the solution of the khrono-metric model \cite{Kovachik:2023fos}. Next, we take $s_S\to\infty$ with the ratio $w\equiv s_V^2/s_S^2$ kept fixed and we show that the result does not depend on $w$. 
Lastly, we discuss corrections to the limiting solution. 

To proceed, we divide Eqs.~(\ref{perteqs}) by $s_V^2$ and group different terms by inverse powers of  $w$ and $s_V^2=ws_S^2$. The result can be cast into the form,
\bseq
\label{perturbed*}
\begin{align}
  \frac{2\mathcal{C}}{\xi^2}-\frac{1}{w}\,\frac{U\mathcal{D}'} {\xi^2} & = \frac{1}{w s_S^2} \Sigma_1 \;, \label{perturbed1}
 \\
  \frac{(U^2\mathcal{C})'}{\xi U}- \frac{1}{w}\,\frac{\mathcal{D}}{\xi^3}  & = \frac{1}{w s_S^2}\Sigma_2\label{perturbed2}\;,
 \end{align}
 \eseq
where we have introduced the following combinations:
\bseq
\label{perturbedS*}
\begin{align}
\mathcal{C} &= - \xi U B_2'- B_1+(\xi U)'B_2 \;,\label{curlyC}\\
\mathcal{D} &= - \xi^2 U B_1'+(2\xi U-\xi^2 U') B_1  - 2\xi B_2\label{curlyD}\;,\\
  \Sigma_1 &=  ( V^2 B_1')' + \left( V V''+\frac{fU''}{U}\right)B_1  +\frac{2U'}{\xi} B_2\;,  \label{perturbedS1}
 \\ 
 \Sigma_2 &=-(V^2 B_2')' - \frac{U'}{\xi} B_1+
  \left(\frac{V^2U''}{U} -\frac{f'}{\xi} + 2 U'^2  +\frac{4UU'}{\xi}\right) B_2   \;. \label{perturbedS2}
 \end{align}
\eseq
Note that the function ${\cal C}(\xi)$ coincides with the twist function $\hat\omega(\xi)$ from Eq.~(\ref{omegalin}).

 \subsection{Limit $s_V \to \infty$, $s_S$ -- fixed}
 \label{LargeRLimit}
 
 Taking the hierarchy of $s_V \gg s_S$ with $s_S$ finite corresponds to the limit  $w\to\infty$. Then Eq.~(\ref{perturbed1}) reduces to ${\cal C}=0$ which gives a simplified differential equation,
\begin{align}
\label{sSLimitEqn1}
- \xi U B_2'-B_1+ (\xi U)' B_2=0~~~~\Longleftrightarrow~~~~
\left( \frac{B_2}{\xi  U}\right)' = - \frac{B_1}{\xi^2 U^2}\;.
\end{align}
Defining $\chi = B_2/\xi U$ we find that the vector field 
is expressed in terms of a 
single scalar function, 
\begin{equation}
B_1 = -\xi^2 U^2 \chi'\;,~~~~~ B_2 = \xi U \chi\;. \label{B12 scalars}
\end{equation}
These expressions match the relations between \ae ther and khronon perturbations in the khrono-metric theory~\cite{Kovachik:2023fos}.\footnote{To establish the map between the khronon perturbation and the functions $B_1$, $B_2$, one has to compare the expressions for the \ae ther components $b^r$, $b_\theta$ from (\ref{bdipole}) with those in \cite{Kovachik:2023fos}, where they are denoted by 
 $\delta u^r$ and $\delta u_\theta$.} 
  To get the equation for $\chi$, we express $\mathcal{C}$ from 
 (\ref{perturbed1}) and insert it into (\ref{perturbed2}). This yields,
\be
\label{newkhronoeq}
(U^3{\cal D}')'  - \frac{2U}{\xi^2 } {\cal D} = \frac{1}{s_S^2}\big(2\xi U \Sigma_2 - (\xi^2 U^2 \Sigma_1)'\big)\;.
\ee
This is a 4th-order equation for the function $\chi(\xi)$. A tedious but straightforward calculation shows that it coincides with Eq.~(4.8) from Ref.~\cite{Kovachik:2023fos}. 
We thus have established that the slowly moving BH solution in the \AE-theory  reduces to that in the khrono-metric theory in the limit $s_V\to\infty$, as expected.

The expressions for the sensitivities in the two theories also match. Indeed, it is shown in \cite{Kovachik:2023fos} that khronon behaves at small $\xi$ as
\be
\chi = \frac{1}{\xi} + \chi_0 + {\cal O}(\xi)\;,
\ee
where the coefficient $\chi_0$ depends on $s_S^2$.
Using this expression in Eqs.~(\ref{B12 scalars}), we have
\be
B_1 = U^2  +\mathcal{O}(\xi^2)=1-\xi+\mathcal{O}(\xi^2)\;,~~~~ 
B_2 = U+\xi U \chi_0+{\cal O}(\xi^2)
=1+\xi\left(\chi_0-\frac{1}{2}\right)+{\cal O}(\xi^2)
\;. \label{B1B2inlargeR}
\ee
Substituting the second expression into
Eq.~(\ref{sigmaInB2}) in the limit $w\to\infty$ we obtain,
\be
\label{senskhron}
\hat\sigma = c_a\left(\frac12 + \frac{s_S^2}{2} + 2s_S^2 \chi_0\right)\;,
\ee
which coincides with Eq.~(5.13) from Ref.~\cite{Kovachik:2023fos}, upon identification of the parameters (\ref{kronocouplings}). 

It has been shown in \cite{Kovachik:2023fos} that the dependence of $\chi_0$ on the khronon speed is described with a good precision by the formula $\chi_0\approx -1/4+0.0244/s_S^2$. This implies that the sensitivity has a finite limiting value (\ref{khronosens}) if we first take $s_V\to\infty$, thereby reducing the \AE-theory to khronon, and then 
also send $s_S\to\infty$. We now demonstrate that the same limiting value is achieved if $s_S$ and $s_V$ are increased uniformly.

\subsection{Limit $s_S\to \infty$, $s_V/s_S$ -- fixed}\label{FiniteRLimit}

When $s_S$ is taken to infinity and $w$ is held fixed, the r.h.s. of Eqs.~(\ref{perturbed*}) vanishes and we obtain a system of two 1st order differential equations for ${\cal C}$ and ${\cal D}$. Combining the two equations, we obtain a homogeneous 2nd order equation for ${\cal D}$,
 \begin{equation}
 (U^3 {\cal D}')' -  \frac{2U}{\xi^2} {\cal D}=0\;.
 \label{D differential}
 \end{equation}
 This has singular points at $\xi=0$ and at merged helicity-0 / helicity-1 / universal horizon $\xi_\star$ where $U(\xi_\star)=0$. Considering the power-law Ansatz ${\cal D}\propto \xi^{n_0}$ for the behavior of the solution in the vicinity of the point $\xi=0$, we obtain two possibilities, $n_0=-1$ or $n_0=2$. However, the definition of ${\cal D}$ in Eq.~(\ref{curlyD}) implies that it must be regular at $\xi\to 0$, so the solution with $\xi^{-1}$ behavior must be rejected. Therefore, only the solution proportional to $\xi^2$ is permissible, implying that ${\cal D}$ vanishes at $\xi=0$. Similar analysis of the power-law behavior ${\cal D}\propto (\xi-\xi_\star)^{n_\star}$ at the universal horizon gives\footnote{One can show \cite{Blas:2011ni,Kovachik:2023fos} that in the limit $s_S\to\infty$ the function $U(\xi)$ for a static Schwarzschild BH takes a simple analytic form, $U(\xi)=\sqrt{1-\xi+27\xi^4/256}$, implying $\xi_\star=4/3$ and $U'(\xi_\star)=-3/2\sqrt{2}$.} $n_\star=-1\pm\sqrt{1+2(\xi_\star U'(\xi_\star))^{-2}}$.
 The divergent solution with the minus sign is inadmissible and must be rejected. The other solution has $n_\star>0$, so ${\cal D}$ also vanishes at $\xi=\xi_\star$.

Thus we have obtained that ${\cal D}$ must vanish at both ends of the interval $[0,\xi_\star]$. We now argue that the only solution of Eq.~(\ref{D differential}) compatible with this requirement is trivial, ${\cal D}(\xi)=0$. Indeed, multiplying this equation by ${\cal D}(\xi)$ and integrating from $\xi=0$ to $\xi_\star$ we obtain, 
 \begin{equation}
 0=\int \dd\xi\, {\cal D}\left((U^3 {\cal D}')'-\frac{2U}{\xi^2}{\cal D}\right)
 =\int \dd\xi\, \left(-U^3 {\cal D}'^2-\frac{2U}{\xi^2}{\cal D}^2\right)\;,
 \end{equation}
where in the second equality we integrated by parts. Since both terms on the r.h.s. are non-positive, the equality can hold only if ${\cal D}$ is identically zero.
It then follows from Eq.~(\ref{perturbed1}) that ${\cal C}(\xi)=0$, and we obtain that the functions $B_1$ and $B_2$ are expressed in terms of khronon, as in Eqs.~(\ref{B12 scalars}). The rest of the analysis of Sec.~\ref{LargeRLimit} follows, implying the limiting value (\ref{khronosens}) for the sensitivity, irrespective of the hierarchy between $s_V$ and $s_S$.

\subsection{Subleading orders}

We have thus far considered two limits, in which the general \ae ther solution reduces to the solution of the khrono-metric theory. We now discuss the corrections responsible for the difference between the two theories. By taking derivatives and linear combinations of Eqs.~(\ref{perturbed*}), they can be brought to the form,
 \begin{align}
  \Delta_\mathcal{C} \mathcal{C}  = \frac{1}{w  s_S^2}\Sigma_\mathcal{C}\;,\qquad 
  %\label{DEforC}
  \Delta_\mathcal{D} \mathcal{D}   = \frac{1}{s_S^2}\Sigma_\mathcal{D}\;,  \label{DEforD}
 \end{align}
where $\Delta_{\mathcal{C},\mathcal{D}}$ are certain 2nd-order differential operators depending only on the background solution $U(\xi)$, whereas the "source terms" $\Sigma_{\mathcal{C},\mathcal{D}}$ contain both the background and perturbation functions $B_1$, $B_2$. The explicit form of the operators and sources is not important for us here; the second equation in (\ref{DEforD}) coincides with  (\ref{newkhronoeq}).

The system (\ref{DEforD}) provides two coupled equations for the functions $B_1(\xi)$, $B_2(\xi)$, which depend on $s_S$ and $w$ as parameters. Recall that 
the background solution $U(\xi)$ depends on $s_S$, but not on $w$. Thus 
the $w$-dependence appears in (\ref{DEforD}) only in the source term of the first equation. We have already seen that in the limit $w\to\infty$ at fixed $s_S$, the solution of Eqs.~(\ref{DEforD}) becomes the khronon solution (\ref{B12 scalars}). Let is denote it by $B_1^{(0)}(\xi;s_S)$, $B_2^{(0)}(\xi;s_S)$, where we explicitly highlight the parametric dependence on $s_S$.
At finite $w$ and at $\xi<\xi_\star$, the solution to Eqs.~(\ref{DEforD}) is organized in 
inverse powers of $ws_S^2$, 
\be
\label{largewgen}
B_{1,2}(\xi;s_S,w) = B_{1,2}^{(0)}(\xi;s_S) + \frac{1}{ws_S^2}
B_{1,2}^{(1)}(\xi;s_S) +\frac{1}{(ws_S^2)^2}
B_{1,2}^{(2)}(\xi;s_S)
+ \cdots\;.
\ee
The situation is more subtle at $\xi=\xi_\star$ since the khronon limiting solution is non-analytic at $\xi_\star$ \cite{Kovachik:2023fos}. This leads to divergence of the sources $\Sigma_{\cal C}$ and $\Sigma_{\cal D}$ at $\xi_\star$, implying that the convergence of the full solution to $B_{1,2}^{(0)}(\xi;s_S)$ at $\xi_\star$ is slower than ${\cal O}\big((ws_S^2)^{-1}\big)$. We do not attempt a detailed study of the expansion at $\xi=\xi_\star$, since we are interested in the sensitivity which is determined by the behavior of the solution at $\xi\to 0$ where the representation (\ref{largewgen}) is valid. 
Substituting it into Eq.~(\ref{sensitivityEqn}) we obtain,\footnote{Equivalent result is reached if we use the alternative representation (\ref{sigmaInB2}).}
\be
\label{sigmalargew}
\frac{\hat\sigma}{c_a} = {B_{1}^{(1)}}'(0;s_S)
+\frac{1}{ws_S^2}
{B_{1}^{(2)}}'(0;s_S)+\ldots\;,
\ee
where we have used the identity ${B_{1}^{(0)}}'(0;s_S)=-1$ which follows from the first equation in~(\ref{B1B2inlargeR}). 

The structure of Eqs.~(\ref{DEforD}) also suggests that at fixed $w$ the solution can be expanded in inverse powers of $s_S^2$. This expansion is admittedly even more subtle than the expansion in $w$ at fixed $s_S$: now also the background solution  $U(\xi;s_S)$ 
depends on $s_S$ and becomes non-differentiable at the universal horizon in the limit $s_S\to \infty$ \cite{Blas:2011ni}. Still, since the universal horizon for any value of $s_S$ lies at $\xi_\star\geq 4/3$ \cite{Kovachik:2023fos}, the expansion is expected to hold at $\xi<4/3$. In particular, it holds in the vicinity of the point $\xi=0$, implying that the individual terms in Eq.~(\ref{sigmalargew}) can  be representated as 
\be
\label{B1assump}
{B_1^{(n)}}'(0;s_S)=\vk^{(n,0)}+\frac{\vk^{(n,1)}}{s_S^2}+\ldots\;,\qquad n=1,2\;.
\ee
In this way we arrive at the expression (\ref{sigmaSeriesDef}) with $\vk^{(1,0)}=(\hat\sigma/c_a)_\infty$, $\vk^{(1,1)}=A_1$, $\vk^{(2,0)}=A_2$\;.
As discussed in Sec.~\ref{sec:sens}, this analytic form is in good agreement with the  exact values of the sensitivity in a wide range of parameters $s_S$ and $s_V$.

\section{Conclusions} 
\label{sec:conc}

In this paper we have found solutions describing slowly moving Schwarzschild BHs in the Einstein--\ae ther theory. The core of our approach was the use of the decoupling limit. Profiting from the fact that the parameters of the \AE-theory have been already constrained to be significantly less than unity by various observations, we neglected the back-reaction of the \ae ther on the metric and reduced the problem to the study of the \ae ther embedding in the fixed Schwarzschild spacetime. 

We worked in the rest frame of the BH where the asymptotic \ae ther field is boosted with respect to the static \ae ther background. We showed that the relevant solution corresponds to a dipolar perturbation of the \ae ther, derived the equations for the radial functions of this perturbation and solved them numerically. The regularity condition at the causal horizons for the helicity-0 and helicity-1 \ae ther waves fix the solution uniquely. The perturbed solution is automatically regular at the Schwarzschild horizon and the universal horizon of the background static BH. For some values of the theory parameters, the solution becomes singular at an inner helicity-1 horizon. This singularity is, however, always shielded from infinity by the outer horizons.

We also found that the universal horizon present in static BHs gets destroyed by the BH motion, except two cases with special choices of the \AE-theory parameters. The first choice is the limit of infinite vector \ae ther waves speed $s_V$, in which case the \AE-theory reduces to the khrono-metric model with hypersurface-orthogonal \ae ther. The second choice corresponds to a special numerical relation between $s_V$ and the speed of scalar \ae ther waves $s_S$. It would be interesting to understand if this latter relation has any other physical implications. We leave this question for future studies.

We matched the asymptotics of the obtained solutions to the \ae ther field produced by a point particle and in this way extracted the effective dimensionless coupling of the BH to the \ae ther $\sigma$ called `sensitivity'. We found it to be proportional to the \AE-theory parameter $c_a$ that has been constrained to be less than $10^{-5}$ by Solar System and pulsar timing analysis. Thus the sensitivity is small.
The proportionality coefficient mildly depends on $s_S$ and $s_V$ and is of order $1$. By studying the limit of the solutions at large values of $s_S$ and $s_V$, we obtained a simple analytic approximation to the sensitivity that describes the results of the exact numerical computation with the accuracy better than $2\%$ in the whole allowed range $s_S,s_V\geq 1$. 

Being dimensionless, the sensitivity is independent of the BH mass. Since dipolar gravitational radiation by a binary system is proportional to the difference in sensitivities of the two companions \cite{Yagi:2013ava}, our results imply that there will be no dipolar emission from Schwarzschild BH binaries. Realistic systems, however, are made of spinning BHs and, since the spin is dimensionless, it is natural to expect that their sensitivities will depend on it. It will be very interesting to apply the methods developed in this work to Kerr BHs in order to find the dependence of the sensitivity on spin. If this happens to be of the same order as the sensitivity itself, it would open an exciting possibility to probe the \AE-theory through the radiation damping of binary BHs using 
combined LISA and LIGO/VIRGO measurements~\cite{Barausse_2016}.

\subsection*{Acknowledgments}
We would like to thank Enrico Barausse, Cliff Burgess, Hari Kunduri, Duncan O'Dell, Pavel Petrov, and Billy Sievers for valuable discussions through the course of this research.
The work of AK and SS is supported by the Natural
Sciences and Engineering Research Council (NSERC) of Canada. The work of AK is also supported by the Ontario Graduate Scholarship (OGS) from the Province of Ontario and McMaster University. Research at Perimeter Institute is
supported in part by the Government of Canada through the Department of Innovation, Science and
Economic Development Canada and by the Province of Ontario through the Ministry of Colleges
and Universities.

%\newpage
%\section*{Appendix}

\appendix
\renewcommand{\thesection}{\Alph{section}}
\renewcommand{\theequation}{\thesection.\arabic{equation}}

%\section{Intuitive Logarithm Terms}\label{sec:logtermneed}
%\section{Solution of the \ae ther equations}

\section{Frobenius analysis of the \ae ther equations}\label{sec:logtermneed}

Here we discuss the Frobenius series of the solutions to the system (\ref{perteqs}) around the singular points $\xi=\{0,\xi_S,\xi_V\}$. This analysis is key for setting 
up the boundary conditions for the numerical solutions. 

The system (\ref{perteqs}) has the schematic form,
\bseq
\label{perteqsApp}
\begin{align}
\label{perteqsApp1}
(p_{11} B_1')'+p_{12}B_2'+q_{11}B_1+q_{12}B_2=0\;,\\
\label{perteqsApp2}
(p_{22} B_2')'+p_{21}B_1'+q_{21}B_1+q_{22}B_2=0\;,
\end{align}
\eseq
where the functions $p_{ij}(\xi)$, $q_{ij}(\xi)$, $i,j=1,2$, are analytic at $\xi>0$. At $\xi=0$, they have poles of first ($p_{12}$ and $p_{21}$) and second ($q_{ij}$) order. The functions $p_{11}$ and $p_{22}$ have single zeros at the scalar and vector horizons, respectively. The system can be compactly written as
\be
\label{perteqsCompact}
{\bf B}''={\bf P}{\bf B}'+{\bf Q}{\bf B}\;,
\ee
using the matrix notations,
\be
\label{Bmatrices}
{\bf B}=\begin{pmatrix}
B_1\\
B_2
\end{pmatrix}\;,~~~~~
{\bf P}=-\begin{pmatrix}
p_{11}'/p_{11}&p_{12}/p_{11}\\
p_{21}/p_{22}&p_{22}'/p_{22}
\end{pmatrix}\;,~~~~~
{\bf Q}=-\begin{pmatrix}
q_{11}/p_{11}&q_{12}/p_{11}\\
q_{21}/p_{22}&q_{22}/p_{22}
\end{pmatrix}\;.
\ee

\subsection{Power series at horizons}

Let us consider the scalar horizon $\xi_S$ and assume $s_S\neq s_V$ implying 
$\xi_S\neq \xi_V$. The matrices ${\bf P}$, ${\bf Q}$ have simple poles at $\xi_S$, so we write
\be
\label{PLaurentS}
{\bf P}(\xi)=\frac{{\bf P}_{(-1)}}{\xi-\xi_S}+{\bf P}_{(0)}+{\bf P}_{(1)} (\xi-\xi_S)+\ldots\;,
\ee
and similarly for ${\bf Q}$. Since we are interested in the solution ${\bf B}(\xi)$ regular at $\xi_S$, we use a Taylor-series Ansatz,
\be
\label{BTaylorS}
{\bf B}(\xi)={\bf B}_{(0)}+{\bf B}_{(1)} (\xi-\xi_S)+{\bf B}_{(2)}(\xi-\xi_S)^2+\ldots\;.
\ee
Substituting it into Eq.~(\ref{perteqsCompact}) and equating coefficients of same powers of $(\xi-\xi_S)$, we get
\bseq
\label{Scoeffeqs}
\begin{align}
\label{Scoeffeqs1}
&0={\bf P}_{(-1)} {\bf B}_{(1)} +{\bf Q}_{(-1)} {\bf B}_{(0)}\;,\\
&(n\!+\!2)(n\!+\!1){\bf B}_{(n+2)}=(n\!+\!2){\bf P}_{(-1)} {\bf B}_{(n+2)}
+\!\!\sum_{m=0}^{n+1}\!\big[(n\!+\!1\!-\!m){\bf P}_{(m)}\!+{\bf Q}_{(m-1)}\big]{\bf B}_{(n+1-m)}\;,\notag\\
&~\qquad\qquad\qquad\qquad\qquad\qquad\qquad\qquad\qquad\qquad\qquad\qquad\qquad\qquad
n=0,1,2,\ldots\;.
\label{Scoeffeqs2}
\end{align}
\eseq
The first equation gives a relation between ${\bf B}_{(0)}$ and ${\bf B}_{(1)}$. Performing the Taylor expansion of functions $p_{ij}(\xi)$, $q_{ij}(\xi)$ similar to (\ref{BTaylorS}) and using $p_{11(0)}=0$, we have
\be
\label{PQsingS}
{\bf P}_{(-1)}=-\begin{pmatrix}
1&p_{12(0)}/p_{11(1)}\\
0&0
\end{pmatrix}\;,~~~~~~
{\bf Q}_{(-1)}=-\begin{pmatrix}
q_{11(0)}/p_{11(1)}&q_{12(0)}/p_{11(1)}\\
0&0
\end{pmatrix}\;,
\ee
so that the relation (\ref{Scoeffeqs1}) takes the form,
\be
\label{SBbc}
p_{11(1)} B_{1(1)}+p_{12(0)}B_{2(1)}+q_{11(0)}B_{1(0)}+q_{12(0)}B_{2(0)}=0\;.
\ee
This can be used to express e.g. $B_{2(1)}$ in terms of three other parameters $B_{1(0)}$, $B_{2(0)}$, $B_{1(1)}$ which remain free. 
The relations (\ref{Scoeffeqs2}) then serve to determine higher-order Taylor coefficients ${\bf B}_{(n+2)}$, $n=0,1,\dots$, in terms these free parameters. Indeed, the determinant of the matrix multiplying ${\bf B}_{(n+2)}$ is 
\be
\det\big[(n+2)((n+1){\bf I}-{\bf P}_{(-1)})]=(n+2)^3(n+1)\neq 0\;,
\ee
and hence this matrix is invertible. Here we have used the notation 
${\bf I}$ for the unit $2\times 2$ matrix. 
tayWe can summarize our results by saying that the regular solution\footnote{It can be shown that the fourth linearly independent solution of the system (\ref{perteqsCompact}) in the vicinity of $\xi_S$ has a logarithmic singularity, ${\bf B}\propto \log(\xi-\xi_S)$.} $B_1(\xi)$, $B_2(\xi)$ at $\xi=\xi_S$ admits a Taylor expansion of the form,
\bseq
\label{BTaylorSfin}
\begin{align}
B_1(\xi)  &= a_{S1}  + a_{S2} (\xi\!-\!\xi_S)+ f_{S(2)}(a_{S1},a_{S2},a_{S3}) (\xi\!-\!\xi_S)^2 + f_{S(3)}(a_{S1},a_{S2},a_{S3}) (\xi\!-\!\xi_S)^3 + \ldots, \\
B_2 (\xi) &= a_{S3}  + g_{S(1)}(a_{S1},a_{S2},a_{S3}) (\xi\!-\!\xi_S)+ g_{S(2)}(a_{S1},a_{S2},a_{S3}) (\xi\!-\!\xi_S)^2 \notag\\
&\qquad\qquad\qquad\qquad\qquad\qquad
\qquad~~~
+ g_{S(3)}(a_{S1},a_{S2},a_{S3}) (\xi\!-\!\xi_S)^3+ \ldots\;,
\end{align}
\eseq
with three free parameters $a_{Si}$, $i=1,2,3$ and known linear functions $f_{S(n)}$, $g_{S(n)}$ of these parameters, whose form is fixed by the Taylor expansion of the coefficients in Eqs.~(\ref{perteqsApp}). 

The analysis at the vector horizon $\xi_V$ proceeds exactly in the same way and leads to the series,
\bseq
\label{BTaylorVfin}
\begin{align}
B_1(\xi)  &= a_{V1} \! + a_{V2} (\xi\!-\!\xi_V)+ \!f_{V(2)}(a_{V1},a_{V2},a_{V3}) (\xi\!-\!\xi_V)^2 + \!f_{V(3)}(a_{V1},a_{V2},a_{V3}) (\xi\!-\!\xi_V)^3 + \ldots, \\
B_2 (\xi) &= a_{V3}  + g_{V(1)}(a_{V1},a_{V2},a_{V3}) (\xi\!-\!\xi_V)+ g_{V(2)}(a_{V1},a_{V2},a_{V3}) (\xi\!-\!\xi_V)^2 \notag\\
&\qquad\qquad\qquad\qquad\qquad\qquad
\qquad~~~~~
+ g_{V(3)}(a_{V1},a_{V2},a_{V3}) (\xi\!-\!\xi_V)^3+ \ldots\;.
\end{align}
\eseq

The situation differs in the special case $s_S=s_V$ implying that the two horizons coincide, $\xi_S=\xi_V$. From the explicit form of Eqs.~(\ref{perteqs}), we see that in this case $p_{11}(\xi)=p_{22}(\xi)$, $p_{12}(\xi)=p_{21}(\xi)=0$, so that the matrix ${\bf P}$ in Eq.~(\ref{perteqsCompact}) becomes proportional to a unit matrix,
\be
\label{Pcoinc}
{\bf P}(\xi)=-\frac{p_{11}'(\xi)}{p_{11}(\xi)}\,{\bf I}\;.
\ee
For a regular solution (\ref{BTaylorS}), one then expresses the coefficient of the linear term through the zeroth-order coefficient,
\be
{\bf B}_{(1)}={\bf Q}_{(-1)} {\bf B}_{(0)}\;.
\ee
All other coefficients are also expressed in terms of ${\bf B}_{(0)}$, which itself remains unconstrained. The general regular solution thus has two free parameters --- the two components of ${\bf B}_{(0)}$.

\subsection{Power series with logs at $\xi=0$}

The analysis of the solution at $\xi=0$ is slightly more complicated. To facilitate it, we once more rewrite the Eq.~(\ref{perteqsCompact}), now in the first-order form. Let us introduce a four-component vector and $4\times 4$ matrix
\be
\label{4notat}
\mathbb{B}(\xi)=\begin{pmatrix}
\xi {\bf B}'(\xi)\\
{\bf B}(\xi)
\end{pmatrix}\;,~~~~~~
\mathbb{P}(\xi)=\begin{pmatrix}
{\bf P}(\xi)+\frac{{\bf I}}{\xi}&\xi{\bf Q}(\xi)\\
\frac{\bf I}{\xi}&0
\end{pmatrix}\;.
\ee
Then (\ref{perteqsCompact}) is equivalent to
\be
\label{4sys}
\mathbb{B}'=\mathbb{P}\, \mathbb{B}\;.
\ee
The matrix $\mathbb{P}$ has a simple pole at $\xi=0$, so we can write,
\be
\label{4Pseries}
\mathbb{P}(\xi)=\frac{\mathbb{P}_{(-1)}}{\xi}+\mathbb{P}_{(0)}+\mathbb{P}_{(1)}\xi+\ldots\;.
\ee
Explicit form of the matrix coefficients $\mathbb{P}_{(n)}$ can be computed from the original Eqs.~(\ref{perteqs}) and the Taylor expansion of the background function, $U(\xi)=1-\xi/2+\ldots$. In particular, the first two coefficients can be found analytically,
\bseq
\label{P-1P0explicit}
\begin{align}
\label{P-1explicit}
&\mathbb{P}_{(-1)}=\begin{pmatrix}
1& -2(1-w)&2(1+w)
&-2(1+w)\\
\frac{1}{w}-1&1&-\frac{2}{w}&\frac{2}{w}\\
1&0&0&0\\
0&1&0&0
\end{pmatrix}\;,\\
\label{P0explicit}
&\mathbb{P}_{(0)}=\begin{pmatrix}
1& -(1-w)&2w
&-\frac{1+s_S^2}{s_S^2}\\
\frac{1-w}{2w}&1&-\frac{1}{2s_V^2}-\frac{3}{2w}+1
&\frac{1}{s_V^2}+\frac{2}{w}-2\\
0&0&0&0\\
0&0&0&0
\end{pmatrix}\;.
\end{align}
\eseq
where we have introduced the notation $w\equiv s_V^2/s_S^2$. Higher-order coefficients can only be found numerically, once the background function $U(\xi)$ has been computed.

We now use the Frobenius Ansatz,
\be
\label{FrobAns}
\mathbb{B}(\xi)=\xi^\kappa \big(\mathbb{B}_{(\kappa,0)}+\mathbb{B}_{(\kappa,1)}\xi+\mathbb{B}_{(\kappa,2)}\xi^2+\ldots\big)\;,
\ee
and find that the necessary condition for it to go through Eq.~(\ref{4sys}) is 
\be
\label{eigeneq}
\kappa \,\mathbb{B}_{(\kappa,0)}=\mathbb{P}_{(-1)}\mathbb{B}_{(\kappa,0)}\;.
\ee
Since $\mathbb{B}_{(\kappa,0)}\neq 0$
by construction, we conclude that $\kappa$ must be an eigenvalue of the matrix $\mathbb{P}_{(-1)}$. Using the explicit form (\ref{P-1explicit}), we find four eigenvalues which are independent of $w$,
\be
\label{kappas}
\kappa=-2,~0,~1,~3\;.
\ee
The solution with $\kappa=-2$ diverges at $\xi=0$ and must be rejected. Other eigenvalues give rise to three linearly independent solutions which are finite at $\xi=0$ and therefore acceptable. The corresponding eigenvectors are
\be
\label{eigenvect}
\mathbb{B}_{(0,0)}=a_{01}\begin{pmatrix}
0\\0\\1\\1
\end{pmatrix}\;,\qquad~~
\mathbb{B}_{(1,0)}=a_{02}\begin{pmatrix}
1\\\frac{1+w}{2}\\1\\\frac{1+w}{2}
\end{pmatrix}\;,\qquad~~
\mathbb{B}_{(3,0)}=a_{03}\begin{pmatrix}
3\\-\frac{3}{2}\\1\\-\frac{1}{2}
\end{pmatrix}\;,
\ee
where $a_{0i}$, $i=1,2,3$, are free parameters.

At the next step, one would like to reconstruct all coefficients in the series (\ref{FrobAns}) order by order. However, the fact that the eigenvalues (\ref{kappas}) are separated by integers leads to an obstruction \cite{bender1999advanced}. To see this, let us pick some $\kappa$ and write the equation for the $n$th coefficient following from (\ref{4sys}),
\be
\label{obstreq}
\mathbb{M}_{(\kappa,n)}\mathbb{B}_{(\kappa,n)}=\mathbb{N}_{(\kappa,n)}\;,~~
\text{where}~~\mathbb{M}_{(\kappa,n)}=(\kappa+n)\mathbb{I}-\mathbb{P}_{(-1)}\;,~~
\mathbb{N}_{(\kappa,n)}=\sum_m^{n-1}\mathbb{P}_{(m)}\mathbb{B}_{(\kappa,n-1-m)}\;, ~~n\geq 1.
\ee
Iff $(\kappa+n)$ is also an eigenvalue of $\mathbb{P}_{(-1)}$, the matrix $\mathbb{M}_{(\kappa,n)}$ is degenerate. A solution $\mathbb{B}_{(\kappa,n)}$ to Eq.~(\ref{obstreq}) then exists only if the vector $\mathbb{N}_{(\kappa,n)}$ belongs to a lower-dimensional subspace spanned by non-null eigenvectors of $\mathbb{M}_{(\kappa,n)}$. Clearly, such situation is not generic.

In the general case, one needs to modify the Ansatz (\ref{FrobAns}) by adding the solution corresponding to the eigenvalue $(\kappa+n)$, multiplied by a logarithm,
\bseq
\label{FrobAnsLog}
\begin{align}
\mathbb{B}(\xi)=&\xi^\kappa \big(\mathbb{B}_{(\kappa,0)}+\mathbb{B}_{(\kappa,1)}\xi+\mathbb{B}_{(\kappa,2)}\xi^2+\ldots\big)
+\log{\xi}\cdot \mathbb{B}_{(\kappa+n)}(\xi)\;,\\
&\text{with}~~\mathbb{B}_{(\kappa+n)}(\xi)=\xi^{\kappa+n}\big(\mathbb{B}_{(\kappa+n,0)}
+\mathbb{B}_{(\kappa+n,1)}\xi+\ldots\big)\;.
\end{align}
\eseq
Substituting this expression into Eq.~(\ref{4sys}) and using that $\mathbb{B}_{(\kappa+n)}(\xi)$ is itself a solution, we see that all terms with $\log \xi$ cancel, whereas the equation for the coefficient $\mathbb{B}_{(\kappa,n)}$ gets modified,
\be
\label{obstreq1}
\mathbb{M}_{(\kappa,n)}\mathbb{B}_{(\kappa,n)}=-\mathbb{B}_{(\kappa+n,0)}+\mathbb{N}_{(\kappa,n)}\;.
\ee
We now recall that $\mathbb{B}_{(\kappa+n,0)}$ is the null vector of $\mathbb{M}_{(\kappa,n)}$. Thus adjusting its normalization, we can cancel any component of $\mathbb{N}_{(\kappa,n)}$ along this direction. As a result, the r.h.s. of (\ref{obstreq1}) will lie in the subspace of the non-null eigenvectors of $\mathbb{M}_{(\kappa,n)}$, admitting a solution for $\mathbb{B}_{(\kappa,n)}$. Note that this solution is not unique: we can add to it any multiple of $\mathbb{B}_{(\kappa+n,0)}$, which just means that we can add a solution corresponding to the eigenvalue $(\kappa+n)$ with an arbitrary coefficient.

Let us now see how the above discussion applies to our particular case. Starting with the solution with $\kappa=0$, we note that the first obstruction can already arise when solving for the coefficient of the linear term $\mathbb{B}_{(0,1)}$. The equation for it reads,
\be
\label{B01eq}
\big(\mathbb{I}-\mathbb{P}_{(-1)}\big)\, \mathbb{B}_{(0,1)}=\mathbb{P}_{(0)}\mathbb{B}_{(0,0)}\;,
\ee
and the matrix $(\mathbb{I}-\mathbb{P}_{(-1)})$ is degenerate.
However, using the explicit expression (\ref{P0explicit}) and the form of $\mathbb{B}_{(0,0)}$ given in Eq.~(\ref{eigenvect}), we find that the r.h.s. is also special, so that the solution actually exists,
\be
\label{B01explicit}
\mathbb{B}_{(0,1)}=a_{01}\begin{pmatrix}
0\\
\frac{w}{2}-\frac{1}{4s_S^2}-\frac{1}{4}\\
0\\
\frac{w}{2}-\frac{1}{4s_S^2}-\frac{1}{4}
\end{pmatrix}+\mathbb{B}_{(1,0)}\;,
\ee
where $\mathbb{B}_{(1,0)}$ is given by the second equation in (\ref{eigenvect}). Thus, no logarithms are required at the linear order.

On the other hand, the numerical calculations show that the obstruction becomes real at the cubic order and the coefficient $\mathbb{B}_{(0,3)}$ cannot be determined without adding logarithmic terms. Returning to the old notations, we arrive at the following series for the three-parameter family of solutions which are finite at $\xi=0$:
\bseq
\label{BTaylor0fin}
\begin{align}
B_1(\xi) & = a_{01}  + a_{02}\, \xi + f_{0(2)}(a_{01}, a_{02})\, \xi^2 + a_{03}\, \xi^3 + f_{0(4)}(a_{01}, a_{02},a_{03}) \,\xi^4+\ldots \notag\\
&\qquad\qquad +\log\xi\cdot\big(h_{0(3)}(a_{01},a_{02})\,\xi^3+h_{0(4)}(a_{01},a_{02})\,\xi^4+\ldots\big)\;,\\
B_2(\xi) & = a_{01}  + g_{0(1)}(a_{01},a_{02})\, \xi + g_{0(2)}(a_{01}, a_{02})\, \xi^2 + g_{0(3)}(a_{01}, a_{02},a_{03}) \,\xi^3+\ldots\notag\\
&\qquad\qquad +\log\xi\cdot\big(k_{0(3)}(a_{01},a_{02})\,\xi^3+k_{0(4)}(a_{01},a_{02})\,\xi^4+\ldots\big)\;,
\end{align}
\eseq
where $a_{01}$, $a_{02}$, $a_{03}$ are free parameters and $f_{0(n)}$, $g_{0(n)}$, $h_{0(n)}$, $k_{0(n)}$ are known linear functions of them.

\section{Numerical method}
\label{allcoefficients}

The numerical solutions of Eqs.~(\ref{perteqs}) presented in the main text are obtained with the following method. Given the numerical background solution $U(\xi)$ for a choice of $s_S$ and fixing $s_V$, we compute the functions $f_{S(n)}$, $g_{S(n)}$, etc. in the series (\ref{BTaylorSfin}), (\ref{BTaylorVfin}), (\ref{BTaylor0fin}) for the solution at the singular points. Let us focus on $\xi_S$ and assume, for concreteness, the ordering $0<\xi_S<\xi_V$ (the case $0<\xi_V<\xi_S$ is treated analogously). We step away a small distance $\pm\epsilon$ (typically, we take $\epsilon\sim 2\times 10^{-3}$) from the point $\xi_S$ and numerically integrate Eqs.~(\ref{perteqs}) with the initial conditions set by the series (\ref{BTaylorSfin}) truncated at cubic order. 
We obtain three basis solutions\footnote{We use vector notations of Appendix~\ref{sec:logtermneed} for the two-component solutions.} ${\bf B}^{(S|i)}(\xi)$, $i=1,2,3$, corresponding to three choices of parameters,
\be
\{a_{S1},a_{S2},a_{S3}\}=
\{1,0,0\}\;,~~
\{0,1,0\}\;,~~
\{0,0,1\}\;.
\ee
The numerical solutions are interpolated over the interval $(\xi_S-\epsilon,\xi_S+\epsilon)$ using the series (\ref{BTaylorSfin}) to obtain the functions regular in the range $(0,\xi_V)$.

The procedure is repeated for the points $\xi=0$ and $\xi_V$, giving in each case a basis of three solutions in the ranges $[0,\xi_S)$ and $(\xi_S,\xi_{\rm max}]$, respectively; here $\xi_{\rm max}>\xi_V$ is the maximal value of the coordinate used in our calculations. 
The global solution we are after is then represented as
\begin{align}
{\bf B}(\xi) =
&\big[a_{01} {\bf B}^{(0|1)}(\xi) + a_{02} {\bf B}^{(0|2)}(\xi) + a_{03} {\bf B}^{(0|3)}(\xi) \big] \Theta(m_1-\xi)\notag\\
+&\big[a_{S1} {\bf B}^{(S|1)}(\xi) + a_{S2} {\bf B}^{(S|2)}(\xi) + a_{S3} {\bf B}^{(S|3)}(\xi) \big] \Theta(\xi-m_1)\Theta(m_2-\xi)\notag\\
+&\big[a_{V1} {\bf B}^{(V|1)}(\xi) + a_{V2} {\bf B}^{(V|2)}(\xi) + a_{V3} {\bf B}^{(V|3)}(\xi) \big] \Theta(\xi-m_2)\;,
\end{align}
where $\Theta(x)$ is the Heaviside function and $m_1$, $m_2$ are the points in the middle of the intervals $(0,\xi_S)$ and $(\xi_S,\xi_V)$. We fix the normalization $a_{01}=1$ and construct a system of linear equations for the rest of the coefficients by imposing the continuity and smoothness of the solution at points $m_1$ and $m_2$,
\bseq
\begin{align}
&\sum_{i=1}^3 a_{0i} {\bf B}^{(0|i)}(m_1)=\sum_{i=1}^3 a_{Si} {\bf B}^{(S|i)}(m_1)\;,\\
&\sum_{i=1}^3 a_{0i} {{\bf B}^{(0|i)}}'(m_1)=\sum_{i=1}^3 a_{Si} {{\bf B}^{(S|i)}}'(m_1)\;,
\end{align}
\eseq
and similarly at $m_2$. This gives us $8$ linear equations for $8$ remaining coefficients $a_{02}$, $a_{03}$, $a_{Si}$, $a_{Vi}$, $i=1,2,3$. Solving this system, we completely determine the solution.

The above procedure was implemented using Mathematica \cite{Mathematica}. 
\bibliographystyle{JHEP}
\bibliography{references.bib}

\end{document}